\documentclass[11pt,a4paper]{article}
\usepackage{jcappub}
\usepackage{lmodern}
\usepackage{graphicx,color}
\graphicspath{{figures/}}
\usepackage{comment}
\usepackage{cancel}

\usepackage[T1]{fontenc}
\usepackage[utf8]{inputenc}
\usepackage{color}
\usepackage{amsmath}
\usepackage{amssymb}
\usepackage{esint}
\usepackage{tikz}           
\usetikzlibrary{positioning}

\makeatletter

\definecolor{somegreen}{cmyk}{0,0.49,0.98,0.09}
\definecolor{red}{rgb}{1,0,0}
\definecolor{magenta}{cmyk}{0,1,0,0}
\definecolor{lavender}{cmyk}{0.16,0.67,0,0.57}
\definecolor{darkgreen}{rgb}{0,0.65,0.05}
\definecolor{antiquefuchsia}{rgb}{0.33, 0.1, 0.89}

\newcommand{\dd}{\textrm{d}}
\newcommand{\bq}{{\bf{q}}}

\let\jnl@style=\rm
\def\ref@jnl#1{{\jnl@style#1}}

\def\aj{\ref@jnl{AJ}}                   
\def\actaa{\ref@jnl{Acta Astron.}}      
\def\araa{\ref@jnl{ARA\&A}}             
\def\apj{\ref@jnl{ApJ}}                 
\def\apjl{\ref@jnl{ApJ}}                
\def\apjs{\ref@jnl{ApJS}}               
\def\ao{\ref@jnl{Appl.~Opt.}}           
\def\apss{\ref@jnl{Ap\&SS}}             
\def\aap{\ref@jnl{A\&A}}                
\def\aapr{\ref@jnl{A\&A~Rev.}}          
\def\aaps{\ref@jnl{A\&AS}}              
\def\azh{\ref@jnl{AZh}}                 
\def\baas{\ref@jnl{BAAS}}               
\def\bac{\ref@jnl{Bull. astr. Inst. Czechosl.}}
\def\caa{\ref@jnl{Chinese Astron. Astrophys.}}
\def\cjaa{\ref@jnl{Chinese J. Astron. Astrophys.}}
\def\icarus{\ref@jnl{Icarus}}           
\def\jcap{\ref@jnl{J. Cosmology Astropart. Phys.}}
\def\jrasc{\ref@jnl{JRASC}}             
\def\memras{\ref@jnl{MmRAS}}            
\def\mnras{\ref@jnl{MNRAS}}             
\def\na{\ref@jnl{New A}}                
\def\nar{\ref@jnl{New A Rev.}}          
\def\pra{\ref@jnl{Phys.~Rev.~A}}        
\def\prb{\ref@jnl{Phys.~Rev.~B}}        
\def\prc{\ref@jnl{Phys.~Rev.~C}}        
\def\prd{\ref@jnl{Phys.~Rev.~D}}        
\def\pre{\ref@jnl{Phys.~Rev.~E}}        
\def\prl{\ref@jnl{Phys.~Rev.~Lett.}}    
\def\pasa{\ref@jnl{PASA}}               
\def\pasp{\ref@jnl{PASP}}               
\def\pasj{\ref@jnl{PASJ}}               
\def\rmxaa{\ref@jnl{Rev. Mexicana Astron. Astrofis.}}%
\def\qjras{\ref@jnl{QJRAS}}             
\def\skytel{\ref@jnl{S\&T}}             
\def\solphys{\ref@jnl{Sol.~Phys.}}      
\def\sovast{\ref@jnl{Soviet~Ast.}}      
\def\ssr{\ref@jnl{Space~Sci.~Rev.}}     
\def\zap{\ref@jnl{ZAp}}                 
\def\nat{\ref@jnl{Nature}}              
\def\iaucirc{\ref@jnl{IAU~Circ.}}       
\def\aplett{\ref@jnl{Astrophys.~Lett.}} 
\def\apspr{\ref@jnl{Astrophys.~Space~Phys.~Res.}}
\def\bain{\ref@jnl{Bull.~Astron.~Inst.~Netherlands}}
\def\fcp{\ref@jnl{Fund.~Cosmic~Phys.}}  
\def\gca{\ref@jnl{Geochim.~Cosmochim.~Acta}}   
\def\grl{\ref@jnl{Geophys.~Res.~Lett.}} 
\def\jcp{\ref@jnl{J.~Chem.~Phys.}}      
\def\jgr{\ref@jnl{J.~Geophys.~Res.}}    
\def\jqsrt{\ref@jnl{J.~Quant.~Spec.~Radiat.~Transf.}}
\def\memsai{\ref@jnl{Mem.~Soc.~Astron.~Italiana}}
\def\nphysa{\ref@jnl{Nucl.~Phys.~A}}   
\def\physrep{\ref@jnl{Phys.~Rep.}}   
\def\physscr{\ref@jnl{Phys.~Scr}}   
\def\planss{\ref@jnl{Planet.~Space~Sci.}}   
\def\procspie{\ref@jnl{Proc.~SPIE}}   

\makeatother

\makeatother

\begin{document}
\title{ Improving precision and accuracy in cosmology with model-independent spectrum and bispectrum}
\author[a]{Luca Amendola,}
\author[b]{Marco Marinucci,}
\author[c]{Massimo Pietroni,}
\author[a,d,e,f]{and Miguel Quartin}
\affiliation[a]{Institute of Theoretical Physics, Philosophenweg 16, Heidelberg University, 69120, Heidelberg, Germany}
\affiliation[b]{Physics department, Technion, 3200003 Haifa, Israel}
\affiliation[c]{Department of Mathematical, Physical and Computer Sciences, University of Parma, and INFN, Gruppo Collegato di Parma, Parco Area delle Scienze 7/A, 43124, Parma, Italy}
\affiliation[d]{Instituto de Física, Universidade Federal do Rio de Janeiro, 21941-972, Rio de Janeiro, RJ, Brazil}
\affiliation[e]{Observatório do Valongo, Universidade Federal do Rio de Janeiro, 20080-090, Rio de Janeiro, RJ, Brazil}
\affiliation[f]{PPGCosmo, Universidade Federal do Espírito Santo, 29075-910, Vitória, ES, Brazil}
\date{\today }
\abstract{
    A new and promising avenue was recently developed for analyzing large-scale structure data with a model-independent approach, in which the linear power spectrum shape is parametrized with a large number of freely varying wavebands rather than by assuming specific cosmological models. We call this method FreePower. 
    Here we show, using a Fisher matrix approach, that precision of this method for the case of the one-loop power spectrum is greatly improved with the inclusion of the tree-level bispectrum. We also show that accuracy can be similarly improved by employing perturbation theory kernels whose structure is entirely determined by symmetries instead of evolution equations valid in particular models (like in the usual Einstein-de Sitter approximation). The main result is that with the Euclid survey one can precisely measure the Hubble function, distance and ($k$-independent) growth rate $f(z)$ in seven redshift bins in the range $z\in [0.6,\, 2.0]$. The typical errors for the lower $z$bins are around 1\% (for $H$), 0.7--1\% (for $D$), and 2--3\% (for $f$).     The use of general perturbation theory  allows us, for the first time, to study constraints on the nonlinear kernels of cosmological perturbations, that is, beyond the linear growth factor, showing that they can be probed at the 10--20\% level. We find that the combination of spectrum and bispectrum is particularly effective in constraining the perturbation parameters, both at linear and quadratic order.    
}
\maketitle

\section{Introduction}

Extracting  information from the abundant new datasets provided by current and future surveys is nowadays one of the most active and important areas of research in cosmology. The main challenge in this endeavour is to improve both in the direction of precision (increase the statistical power) and of accuracy (keep systematics under control and increase the robustness of the results). The goal of this paper is to advance in both directions. We improve the statistical power by making use of non-linear (one-loop) power spectrum and of tree-level bispectrum. We also achieve a high level of model independence by choosing a number of linear power spectrum parameters equal to the number of available $k$-bins, instead of parametrizing the spectrum with a given model or set of templates. We now discuss in turn these two complementary aspects.

One of the most successful theoretical model for the description of the large-scale structure (LSS) is the effective field theory of large scale structure (EFTofLSS ~\cite{Baumann:2010tm,Carrasco:2012cv}, see also \cite{Pietroni:2011iz,Manzotti:2014loa} for a related approach)\footnote{For an alternative approach based on Kinetic Field Theory, see e.g.~\cite{Lilow:2018ejs}.} which provides a rigorous method for the calculation of cosmological correlators up to mildly non-linear scales, including corrections coming from small scale non-linearities. EFT-based models for the power spectrum at one loop have been recently used for the analysis of the BOSS dataset~\cite{Alam:2016hwk}, giving constraints on cosmological parameters that are competitive with CMB based observations~\cite{DAmico:2019fhj,Ivanov:2019pdj, Troster:2019ean, Semenaite:2021aen, Chen:2021wdi}. The results of these and following data analysis (constraints on neutrino masses~\cite{Ivanov:2019hqk,Colas:2019ret}, $H_0$ tension~\cite{Philcox:2020vvt,DAmico:2020ods}, beyond-$\Lambda$CDM model~\cite{Ivanov:2020ril,DAmico:2020kxu,DAmico:2020tty,Piga:2022mge}, redshift space distortions~\cite{Ivanov:2021fbu, DAmico:2021ymi}) represent a great step forward in the study of LSS within the EFTofLSS framework. Moreover, in the BOSS collaboration the bispectrum has been measured and analyses have been performed~\cite{Gil-Marin:2014sta,Gil-Marin:2014baa,Gil-Marin:2016wya}, while the first detection of the BAO features in the bispectrum has been claimed in~\cite{Pearson:2017wtw}. The first analysis using the monopole of the BOSS bispectrum was performed in~\cite{Philcox:2021kcw}, which showed that, for the BOSS volumes and galaxy number densities, the inclusion of the bispectrum monopole has the main effect of reducing the errorbar on $\sigma_8$ of about 13\%, while  the constraints on higher order biases are much strengthened. This was confirmed by the analysis of~\cite{DAmico:2022osl}, which included the one-loop bispectrum monopole and the tree-level bispectrum quadrupole, for which the improvement on the matter density amplitude is about $\sim 30 \%$ and there is minor improvement on the other cosmological parameters. Adding higher multipoles in the BOSS analysis does not seem to have a large impact, see~\cite{Ivanov:2023qzb}.

More recent studies on numerical simulations and forecast for future galaxy surveys have shown that the use of the bispectrum will remarkably improve the constraints on higher order biases. Ref.~\cite{Agarwal:2020lov} showed that by performing a joint analysis of the spectrum and the bispectrum one could reach a $10\%$ accuracy on $f$ and $b_1$ for a Euclid-like survey including also selection effects, while~\cite{Yankelevich:2018uaz} has shown that using only the bispectrum monopole reduces significantly the information content of the bispectrum, a result that was confirmed on simulations~\cite{Pardede:2022udo}. Ref.~\cite{Gualdi:2020ymf} performed a Fisher forecast to study the information content of $m\neq 0$ multipole momenta for the $B^{(l,m)}$ bispectrum, showing an improvement of $\sim 30 \%$ on error bars when higher bispectrum momenta (up to $l=4$) are included in the analysis. A detailed analysis performed on the perturbation theory (PT) challenge simulations~\cite{Nishimichi:2020tvu} using the one-loop bispectrum~\cite{Philcox:2022frc} has shown that going to higher perturbative order enhances the maximum wavevector for the bispectrum, $k_{\rm max}^B$, from $0.06\,h$  Mpc$^{-1}$ to $0.15\,h$ Mpc$^{-1}$ and yields only a slight improvement on $\sigma_8$ and the bias parameters $b_1$, $b_2$ and $b_{\mathcal{G}_2}$ compared to the analysis with the tree-level bispectrum. For a BOSS-like volume, Ref. ~\cite{Philcox:2022frc} found that going to 1-loop for the bispectrum causes an improvement of $\sim 10 \%$ on the galaxy biases.

 Some of these works however have some limitations with respect to the present study. For instance, in~\cite{Yankelevich:2018uaz,Gualdi:2020ymf}, only the linear spectrum, together with the bispectrum, was considered. In~\cite{Agarwal:2020lov} one-loop $P(k)$ and tree-level bispectrum for Euclid was studied, but only at a single redshift.
Perhaps the most important difference, however, is that all the previous analyses were based on specific cosmological models. In fact, the standard approach with analysing LSS in Fourier space has been to assume a given cosmological model (usually $\Lambda$CDM) and use the power spectrum and possibly the bispectrum to fit the model parameters by comparing data and results from Boltzmann codes. 
In particular, the ${\alpha_\parallel,\alpha_\perp,f\sigma_8}$ parametrization is often used in order to improve model-independence (see e.g.~\cite{Zhao:2015gua, BOSS:2016wmc, BOSS:2016hvq, Foroozan:2021zzu}). This procedure, however, still relies on assumptions regarding the shape of $P(k)$.

The challenges of performing model-independent LSS measurements have been discussed  over a decade ago~\cite{Samushia:2010ki}. One proposed approach to obtain some level of model-independence in full-shape power spectrum measurements is to use a set of templates parametrized by as few as one extra single parameter~\cite{Brieden:2021edu}. Recently, further progress has been made and it was shown on~\cite{Amendola:2019lvy} that allowing a completely free $P(k)$ shape, by simply using each $k$-bin as an independent parameter, could result in precise constraints in the Hubble function at low redshift if future galaxy in linear scales and supernova data are analyzed jointly. The effects of the first non-linear corrections in the form of the one-loop EFT terms on this methodology was explored in~\cite{Amendola:2022vte}, where  higher redshifts were analysed without the use of supernovae. It was shown in that work that even allowing a complete general scale-dependent growth $f(k,z)$ and uninformative priors for all bias parameters  one could get $\sim 10\%$ constraints in $H(z)$ in different redshift bins. We call this method FreePower.

The main qualitative difference between the FreePower method and the standard full-shape approach is that in the latter the cosmological parameters enter both in the shape and in the Alcock-Paczyński (AP) and redshift space distortion (RSD) terms. On the other hand, in the FreePower method we decouple the shape, and thus the early universe information, from the one coming from AP, RSD, and the nonlinear evolution of cosmological perturbations. In this way, resulting measurements become independent from the early universe and thus mostly independent from the ones from the CMB. This can be an asset, say, in probing the origins of cosmological tensions. Now, clearly,  AP, RSD effects, and nonlinearities  are also present in the full-shape approach. Thus, in quantitative terms, by modelling the early universe it is possible to extract extra information for a given choice of $k_{\rm max}$ and for the same summary statistics at the price of losing model-independence and a clearer separation between early and late-time constraints.

In this work we include in the FreePower method the one-loop non-linear power spectrum and the tree-level bispectrum and forecast constraints for  Euclid in the redshift range 0.6--2.0. We also take a further step towards model-independence by  including for the first time PT kernels that do not assume a specific cosmology (e.g.~EdS or $\Lambda$CDM) but whose  structure  is dictated only by symmetries, namely, the Extended Galileian Invariance and momentum conservation, see~\cite{DAmico:2021rdb}. This approach requires adding four additional redshift-dependent free parameters to the non-linear kernels, the so called \textit{bootstrap} parameters. 
We find results in agreement with previous studies on the biases  and we also put, for the first time, constraints on  bootstrap parameters. 

With respect to~\cite{Amendola:2022vte} there are several improvements. Beside the major ones, to wit the addition of the bispectrum and of the beyond-EdS bootstrap parameters, we also introduce several technical improvements. First, we include now the IR resummation according to the prescription in~\cite{Ivanov:2019pdj}; second, we  vary separately the angular diameter distance along with the Hubble function; third, we include the Alcock-Paczyński  prefactor that multiplies the power spectrum.

\section{Theoretical power spectrum and bispectrum}
\label{app:spe-bispe}

We model the galaxy power spectrum  $P_{gg}$ according to one-loop perturbation theory plus UV counterterms, shot noise terms, and a smoothing factor that models pairwise velocity smoothing and spectroscopic errors \cite{Ivanov:2019pdj,DAmico:2019fhj}: 
\begin{align}
    P_{gg}(k,\mu,z) & =S_{{\rm g}}(k,\mu,z)^{2}\left[P^{{\rm lin}}(k,\mu,z)+P^{{\rm 1loop}}(k,\mu,z)+P^{{\rm UV}}(k,\mu,z)\right]+P^{{\rm SN}}(z)\,,\label{Pgg}
\end{align}
where $k=(k_{\parallel}^{2}+k_{\perp}^{2})^{1/2}$, $\mu\equiv k_{\parallel}/k$, and $k_{\parallel}$ ($k_{\perp}$) is the component of the wavevector parallel (perpendicular) to the line of sight. The linear contribution is given by 
\begin{equation}
    P^{{\rm lin}}(k,\mu,z)=Z_1({\mathbf{k}};z)^2 G(z)^2 P_1(k)\,,\label{plin}
\end{equation}
where $P_1(k)$ is the linear matter power spectrum  in real space evaluated at $z=z_1$, the lowest redshift value we will consider in the analysis, and $G(z)$ the linear growth factor, normalized as $G(z_1)=1$. The redshift distortion factor is $Z_1=b_1+f\mu^2$, where $b_{1}(z)$ is the linear bias parameter and $f(z)\equiv\dd\log G/\dd\log a$ is the linear growth rate. In \cite{Amendola:2022vte} we considered a $k$-dependent growth rate, but here we decided for simplicity (and for consistency with the form of the PT kernels, see below) to consider $f$ dependent only on $z$. We leave the more general case to future work. 

The complete expressions for $P^{{\rm 1loop}}(k,\mu,z)$
and $P^{\rm UV}$ are given in Appendix~A of \cite{Amendola:2022vte}  (see also~\cite{Ivanov:2019pdj,DAmico:2019fhj}). As anticipated, with respect to that paper, here we take a further step towards model independence, and go beyond the usual EdS approximation for the perturbation theory kernels $Z_{1,2,3}$. We use the expressions derived in  ref.~\cite{DAmico:2021rdb} in the so-called ``bootstrap approach'', which consists in implementing symmetry constraints (from rotational invariance, Extended Galileian Invariance and momentum conservation) instead of computing the kernels by solving the continuity, Euler, and Poisson equations in a fixed cosmological model. This approach is related to an earlier one  based on consistency relations derived in~\cite{Fujita:2020xtd}, which dealt with biased tracers in real space. Ref.~\cite{DAmico:2021rdb}, extended this approach to redshift space, which is essential to derive model-independent cosmological information, and, moreover, highlighted new constraints on the structure of the kernels (see Section 4 of~\cite{DAmico:2021rdb}  for a detailed discussion on the relation between the two approaches).

The explicit expressions of the bootstrap kernels are given in Appendix \ref{app:beds}. They contain four tracer-independent parameters that encode cosmological information at the perturbation level. Besides $f$, which enters at the linear level, there is one parameter entering at the second order,
\begin{equation}
    d_\gamma^{(2)}(z)\,,
\end{equation}
and two parameters entering at third order, 
\begin{equation}
    a_\gamma^{(2)}(z)\,,\;d_{\gamma a}^{(3)}(z)\,.
\end{equation} 
As mentioned, in a given cosmological model the above coefficients can be computed at any redshift by solving the corresponding continuity-Euler-Poisson system. In $\Lambda$CDM they are usually approximated by their EdS values \cite{Bernardeau:2001qr}, listed in (\ref{EdSvals}), which turns out to work at the subpercent level on the scales of interest \cite{Pietroni:2008jx, Donath:2020abv}. In theories beyond $\Lambda$CDM, on the other hand, the effect can be larger, expecially in models featuring nonlinear `screening'  of extra degrees of freedom   \cite{Bose:2018orj, Piga:2022mge}. 

Together with the background quantities, namely the Hubble function $H(z)$ and the angular diameter distance $D(z)$, the parameters $f(z)$, $d_\gamma^{(2)}(z)$, $a_\gamma^{(2)}(z)$, and $d_{\gamma a}^{(3)}(z)$  form the complete set of cosmology-dependent and tracer independent parameters entering our model for the power spectrum and the bispectrum.

Moreover, as in $\Lambda$CDM, the kernels contain four tracer-dependent `bias' parameters,
\begin{equation}
    b_1,\,b_2,\,c_\gamma^{(2)},\,a_{\gamma a}^{(3)}\,,
\end{equation}
where the relation of the last two with the parameters $b_{{\cal G}_2}$ and $b_{{\Gamma}_3}$ used, for instance, in \cite{Ivanov:2019pdj}, is given in (\ref{biasvoc2}).

$P^{\rm UV}$ contains also three `counterterm' parameters, $c_0$, $c_2$, and $\tilde c$, but here we consider only $c_0$ and set the other two to zero, since we model the small-scale bulk flow damping with the overall function $S_g$. We set therefore\footnote{Notice that our $c_0$ corresponds to $c_0^2$ in Ref. \cite{Ivanov:2019pdj}. In App.~A of ref.~\cite{Amendola:2022vte} there are some typos: the factors $c_0^2,c_2^2,\tilde c^2$ should actually be $c_0,c_2,\tilde c$, respectively. These typos did not propagate to the calculation.}
\begin{equation}
    P^{\rm UV}
    =-2G^2 P_1(k)c_{0}k^{2}
    \,.\label{eq:UV-2}
\end{equation}
Since in FreePower we use $P_1(k)$  in wavebands as parameters, we are largely insensitive to the exact location
of the BAO wiggles, 
but there is still sensitivity to the change of slope near the wiggles as the one induced by bulk flows. This can be modelled as  a damping factor on the oscillating
part of the power spectrum (sometimes called IR resummation). We follow the treatment of~\cite{Ivanov:2019pdj} by splitting the linear power spectrum into a smooth and a wiggles-only component. This correction is fixed, that is, we do not vary the parameters inside the correction itself. The smooth spectrum is obtained from the Eisenstein and Hu fitting function for the broadband linear spectrum \cite{Eisenstein:1997jh}, and the wiggles-only spectrum by subtracting the smooth spectrum from the full one. The oscillation scale is taken as $q_{\rm osc}=(2\pi/110)\; h/$Mpc and the cut-off scale as $0.2 h/$Mpc, again as in \cite{Ivanov:2019pdj}. The IR resummation, however, has  an impact of less than 10\% on the final constraints. 

The spectroscopic redshift errors and the Finger-of-God (FoG) effect are modelled through an overall smoothing factor $S_{{\rm g}}(k,\mu,z)^{2}$, described below. Following~\cite{2020A&A...642A.191E,BOSS:2016psr, Amendola:2022vte},
we thus write  
\begin{equation}
    S_{{\rm g}}(k,\mu,z) = \exp\left[-\frac{1}{2}(k\mu\sigma_{{\rm z}})^{2}\right] \, \exp\left[-\frac{1}{2}(k\mu\sigma_{f})^{2}\right],\label{eq:FoG}
\end{equation}
where
\begin{equation}
    \sigma_{{\rm z}}=\sigma_{0}(1+z)H(z)^{-1}\,.
\end{equation}
For an Euclid-like survey, $\sigma_{0}=0.001$ (which, at $z\approx1$, corresponds to an effective smoothing over scales of $\approx10$ Mpc$/h$).
The FoG smoothing $\sigma_f$ is left as a free parameter in each $z$-bin.
We do not allow the AP effect to vary $k,\mu$ inside $S_g$.

Finally, we model the shot-noise spectrum as
\begin{equation}
    P^{{\rm SN}}(z)=\frac{1}{n(z)}(1+P_{{\rm sn}})\,,
\end{equation}
where $n(z)$ is the density of the considered galaxies at redshift $z$, and $P_{{\rm sn}}$ an additional nuisance parameter to account for deviations from the pure Poisson noise. Such a degree of freedom is often used, among other things, to account for effects such as exclusion in the galaxy data~\cite{2013PhRvD..88h3507B}, or fiber-collision effects~\cite{Gil-Marin:2015sqa}. Additional nuisance parameters accounting for scale-dependent stochasticity are sometimes employed, but they can be very degenerate with some counterterm parameters, and some results indicate that their contribution to the final results is small~\cite{Chudaykin:2020aoj}.  Following~\cite{Philcox:2021kcw,Yankelevich:2018uaz}, for the bispectrum, we use  two shot noise nuisance parameters $B_{\rm sn(1)}$ and $B_{\rm sn(2)}$, modelled as 
\begin{equation}
    B^{\rm SN}(\mathbf{k}_{1},\mathbf{k}_{2},\mathbf{k}_{3}) = \frac{1}{n(z)} \Big(P^{{\rm lin}}(\mathbf{k}_{1})+P^{{\rm lin}}(\mathbf{k}_{2}) + P^{{\rm lin}}(\mathbf{k}_{3})\Big) \Big(1+B_{\rm sn(1)}\Big) + \frac{1}{n(z)^2}\Big(1+B_{\rm sn(2)}\Big)\,.
\end{equation}

The vectors $\mathbf{k}$ are defined by $k$ and $\mu$, since $P_{gg}$ does not depend on the azimuthal angle. We adopt throughout bins of width $\Delta k=0.01$ with $k_{\mathrm{min}}=0.01 h$/Mpc,\footnote{We will discuss this choice of $k_{\rm min}$ in Section~\ref{FMs}.} and $\Delta\mu=0.1$, including for the bispectrum. We call every distinct
configuration of $\{k, \mu \}$ for $P$ a vector (also called a wedge), and we associate a Latin index $i$ or $j$ to each vector bin. There are therefore $(k_{\mathrm{max}}/\Delta k) \times(1/\Delta\mu)$ vector bins (due to the symmetry in $\mu$, we need to run $\mu$ only in the range $[0,1]$). We tested that results do not change by more than 5\% degrading the $k$-resolution to  $\Delta k=0.02$, nor by increasing the $\mu$-resolution to 0.05. A degradation in angular resolution to 0.2, on the other hand, induced changes up to 30\%.  All the momentum integrals are performed by integrating power spectra generated with the CLASS Boltzmann code~\cite{Blas:2011rf}. The explicit dependence on $z$ is from now on understood and will be omitted unless necessary for clarity.

The covariance matrix for the power spectrum is 
\begin{equation}
    C_{PP,ij}=2P_{gg}(\mathbf{k}_{i})P_{gg}(\mathbf{k}_{j})\delta_{ij}/N_{P}\,,
\end{equation}
where
\begin{equation}
    N_{P}=\frac{V}{(2\pi)^{2}}k_{i}^{2}\Delta k\Delta\mu
\end{equation}
is the number of vectors per bin.
The power spectrum is assumed to be Gaussian distributed and with a diagonal covariance, but the density contrast is non Gaussian, as the diagonal components are proportional to the square of the nonlinear PS, not the linear one. However, other non-Gaussian contributions, both in the diagonal and the  non diagonal terms, are induced by the trispectrum, the survey geometry, and super sample variance. Analytical covariances have recently been studied in great detail in \cite{Wadekar:2019rdu} and \cite{Wadekar:2020hax}, where it was found that the effect of these additional non-Gaussian terms on parameter constraints, after marginalization over cosmological and bias parameters, is typically less than 10 \% (see for instance Fig. 1 in \cite{Wadekar:2020hax}).

We take the tree-level bispectrum as~\cite{Scoccimarro:1999ed, Desjacques:2016bnm, Yankelevich:2018uaz}
\begin{equation}
    B(\mathbf{k}_{1},\mathbf{k}_{2},\mathbf{k}_{3}) = 2\big[Z_{1}(\mathbf{k}_{1})Z_{1}(\mathbf{k}_{2})Z_{2}(\mathbf{k}_{1},\mathbf{k}_{2})G^4 P_1(k_1)P_1(k_2)+2\;{\rm cycl.}\big] 
    + B^{\rm SN} (\mathbf{k}_{1},\mathbf{k}_{2},\mathbf{k}_{3}) \label{eq:bisp}
\end{equation}
with $\mathbf{k}_{3}=-\mathbf{k}_{1}-\mathbf{k}_{2}$. The triangle configurations are defined by five quantities: $k_{1}$, 
$k_{2}$, $k_{3}$, and the  cosine angles  $\mu_{1}$ and $\mu_{2}$ between the line of sight and ${\mathbf k}_1$ and ${\mathbf k}_2$, respectively. We call every distinct set of these five values a triangle, and we associate to each triangle bin a Latin index $i$ or $j$. The cosine angle $\mu_{3}$ amounts to 
\begin{equation}
    \mu_{3}=-\frac{1}{k_3}(\mu_{1}k_{1}+\mu_{2}k_{2})\,,
\end{equation}
and of course we need to express also the azimuthal angles $\phi_{1},\phi_{3}$ in terms of our five variables (we take $\phi_{2}=0$ without loss of generality).

In the Gaussian approximation, the correlation between triangles $i,j$ can be simplified to (see e.g. \cite{Fry:1992ki,PhysRevD.69.103513,Yankelevich:2018uaz})
\begin{equation}
    C_{BB,ij}=s_{B}\frac{V}{N_{B}}G^6 P_1(k_{1})P_1(k_{2})P_1(k_{3})\delta_{ij}L_{i}\label{eq:cbb}\,,
\end{equation}
where $s_{B}=6,2,1$ for equilateral, isosceles, and scalene triangles, respectively. Also, $L_{i}=2$ for co-linear triangles (i.e. such that $k_{3}=k_{1}+k_{2}$ or $k_{3}=|k_{2}-k_{1}|$)
and 1 otherwise \cite{Chan:2016ehg}. The number of triangles per bin is given by
\begin{equation}
    N_{B}=2\frac{V^{2}}{8\pi^{4}}k_{i_1}k_{i_2}k_{i_3}(\Delta k)^{3}\Sigma(\Omega)\Delta\Omega\,,
\end{equation}
where $k_{i_{1,2,3}}$ are the central value of the $k$-bins, and $\Sigma(\Omega)\Delta\Omega$ is the number of triangles within the angle orientation $\Delta\Omega=(\Delta\mu)^{2}$,
that depends on which coordinate system is used.  For our choice of angular coordinates, one has~\cite{Yankelevich:2018uaz}
\begin{equation}
    \Sigma=\bigg[2\pi\sqrt{1-m_{12}^{2}-\mu_{1}^{2}-\mu_{2}^{2}+2\mu_{1}\mu_{2}m_{12}}\bigg]^{-1},
\end{equation}
where $m_{12}$ is the cosine angle between $\mathbf{k}_{1},\mathbf{k}_{2}$, i.e.,
\begin{equation}
    m_{12} = \frac{\mathbf{k}_{1}\cdot\mathbf{k}_{2}}{k_{1}k_{2}}=\frac{k_{3}^{2}-k_{1}^{2}-k_{2}^{2}}{2k_{1}k_{2}}\,.
\end{equation}
We obviously require  $\mu_{3}\in[-1,1]$ and $k_{3}\in(k_{1}+k_{2},|k_{1}-k_{2}|)$, so only configurations that satisfy these constraints are considered.

The density $\Sigma(\Omega)$ is to be normalized so that $\int\Sigma(\Omega)\Delta\Omega=1$, but of course for finite bins the integral becomes a sum and the normalization  is only approximately true. We therefore renormalize $\Sigma(\Omega)$ so that for every $k_1,k_2,k_3$ the sum $\sum_{ij}\Sigma(\mu_i,\mu_j)\Delta\mu_i\Delta\mu_j$ equals unity. This, however, has only a minor impact on the results. Note that in $N_{B}$ there is an extra factor of two because the bispectrum is symmetric under simultaneous change of sign of both $\mu_{1},\mu_{2}$, so we vary $\mu_{1}$ in $[-1,1]$ but $\mu_{2}$ only between 0 and 1. In fact, since we assume central bin values and a step $\Delta\mu$, we vary $\mu_{1}\in(-1+\Delta\mu/2,1-\Delta\mu/2)$ and $\mu_{2}\in(\Delta\mu/2,1-\Delta\mu/2)$. 

If a vector $i$ of the power spectrum coincides with a side $j_{1},j_{2},j_{3}$ of a bispectrum triangle,  a  cross-correlation arises. We dub this the $PB$ covariance, and it is given by \cite{Yankelevich:2018uaz} 
\begin{equation}
    C_{PB,ij}=2s_{PB}\frac{P_{i}B_{j}}{N_{P}}(\delta_{ij_{1}}+\delta_{ij_{2}}+\delta_{ij_{3}})\,,
\end{equation}
where $s_{PB}=3,2,1$ for equilateral, isosceles, and scalene triangles. We tested that the results change by no more than a few percent on average when the cross-covariance $C_{PB}$ is included, since only a small fraction of triangles has a side in common with a vector, so from now on we neglect it (see also \cite{Yankelevich:2018uaz}). This vastly simplifies the problem because then all covariance matrices are diagonal and inversion does not present any numerical problem.

\section{Fisher matrix}
\label{FMs}

From now on, for conciseness, we refer to the power spectrum and the
bispectrum as $P$ and $B$, respectively. The complete Fisher matrix (FM) $\mathbf{F}$ is given by 
\begin{equation}
    F_{\alpha\beta}=X_{i,\alpha}C_{ij}^{-1}X_{j,\beta}
\end{equation}
 (sum over $i,j$) where $X_{i}=\{P,B\}$ represents the  values of $P$
for every vector bin and of $B$ for every triangle bin, and $()_{,\alpha}$ represents the partial derivative with respect to parameter $\alpha$ calculated on the fiducial. The correlation matrix $C_{ij}$ is formed by the two independent diagonal blocks $C_{PP,ij}$ and $C_{BB,ij}$. Whenever we mention results for $P$ (or $B$) only, we refer obviously to a FM in which $X$ and $C_{ij}$ only contain $P$ (or $B$, respectively) elements. Priors are applied only once to the final FM in all cases.

Vectors, triangles, and volumes are obtained by converting angles
and redshifts to cosine $\mu_r$  and wavenumber $k_r$ by assuming a reference model (subscript $r$). For
any other cosmology, the correction, known as the Alcock-Paczyński
effect, is given by $\mu=\mu_{r}h /\alpha$ and 
$k=\alpha k_{r}$, where $h=H/H_r$ and ~\cite{2000ApJ...528...30M,Amendola:2004be,Samushia:2010ki}
\begin{equation}
\alpha\,=\,\frac{1}{d}\sqrt{\mu_{r}^{2}(h^{2}d^2-1)+1}\,,\label{eq:alpha}
\end{equation}
where $d=D/D_r$. Moreover, all observed spectra get multiplied by a volume-correcting factor $\Upsilon$~\cite{1996MNRAS.282..877B,Seo:2003pu,Quartin:2021dmr}, so that $P_{{\rm gg,\,obs}}\rightarrow{\Upsilon P_{{\rm gg}}}$, where 
\begin{equation}
    \Upsilon=\frac{h}{d^{2}}\,.
\end{equation}
The $\Upsilon$ factor is unnecessary for the tree-level bispectrum since it is fully degenerate with the power spectrum wavebands. 
As anticipated, instead of the usual cosmological parameters, we take the $k$ wavebands of the linear power spectrum as  parameters to be varied in the Fisher matrix. Since we assume a $k$-independent growth rate $f$, spectra at different redshifts scale by the $k$-independent growth $G=\exp\int fd\log a$. Therefore, the free spectral parameters can be taken as $f(z)$ and $P_1=P(k,z=z_1)$ where $z_1$ is the first $z$ bin. Beside the wavebands of $P_1$,
the list of parameters we vary independently at each redshift is then
\begin{equation}
    \theta_{\alpha}=\{\log f,\log H,\log D,\log b_{1},b_{2}, \log c_\gamma^{(2)}, a_{\gamma a}^{(3)}, c_{0},\log\sigma_{f},P_{\mathrm{sn}},B_{\mathrm{sn}(1)},B_{\mathrm{sn}(2)}\}\,,
\end{equation}
plus the three tracer-independent but cosmology-dependent parameters appearing in the beyond-EdS kernels at second and third order\footnote{As discussed in Appendix \ref{app:beds}, the parameter $a_{\gamma a}^{(3)}$ is effectively tracer-dependent in our analysis, as we fix the bias parameter $b_{\Gamma_3}$, see eqs.~(\ref{biasvoc}), (\ref{biasvoc2}).},
\begin{equation}
    \{\log d_{\gamma}^{(2)},\,a_{\gamma}^{(2)},\, d_{\gamma a}^{(3)}\}\,,
\label{beds23}
\end{equation}
for a total of 15 parameters (that we denote collectively as the $z$-dependent parameters) plus a variable number of $\log P_1(k)$
parameters equal to $k_{\mathrm{max}}/\Delta k$. Some
parameters are taken as logarithms so as to enforce positive definiteness and to obtain directly relative errors. The derivatives of the power spectrum $P_{gg}$ with respect to $P_1(k),\beta,b_{1},b_{2}, c_\gamma^{(2)},c_{0}$, are detailed in the Appendix A of \cite{Amendola:2022vte}. Here we convert the derivatives
with respect to $\beta,b_{1}$ into derivatives with respect to $f,b_{1}$, and add derivatives
with respect to the cosmology-dependent parameters of eq.~(\ref{beds23}), which do not present any difficulty. The derivatives with respect to $\sigma_{f}$ and $P_{\mathrm{sn}}$ are also trivial. For the $H$- and $D$-derivatives   which enter through the AP effect in $\mathbf{k}$ (and for the bispectrum,  $\mathbf{k}_{1},\mathbf{k}_{2},\mathbf{k}_{3}$), we adopt a five-point stencil numerical derivative with relative step 0.01. We experimented with other steps and found only very minor differences.

The power spectrum depends on the growth rate $f$ also through the growth
function $G(z)=\exp\big({-\int_{z_1}^{z}f(z')/(1+z') dz'}\big).$ Collecting the terms $P_{(m)}$ that depend on the same power $m$ of the linear spectrum, we can write 
\begin{equation}
    P_{gg}=G^{2}P_{(1)}+G^{4}P_{(2)}\,.
\end{equation}
We need then to differentiate $G(z_j)$ with respect to $f(z_i)$. We have
\begin{equation}
    \frac{\partial G(z_j)}{\partial\log f(z_i)}\approx-f(z_i)G(z_j)\frac{\Delta z}{1+z_i}\,,\qquad (j\ge i >1)\,,
\end{equation}
where $\Delta z=0.2$ is the bin's width and therefore when we take $\partial P_{gg}/\partial\log f$, we need to
include these additional derivatives, 
\begin{equation}
    \frac{\partial P_{gg}(z_j)}{\partial\log f(z_i)}=\cdots  -2f(z_i)G(z_j)^{2}\frac{\Delta z}{1+z_i}P_{(1)}(z_j)-4f(z_i)G(z_j)^{4}\frac{\Delta z}{1+z_i}P_{(2)}(z_j)
    \,,
    \label{eq:offdiagP}
\end{equation}
    where dots indicate contributions from derivatives of the PT kernels.
Similarly, since $B\propto G^{4}$, we have the additional term
\begin{equation}
    \frac{\partial B(z_j)}{\partial\log f(z_i)}= \cdots-4f(z_i)\frac{\Delta z}{1+z_i}B(z_j)\,,\qquad (j\ge i >1)\,.\label{eq:offdiagB}
\end{equation}

One has to be careful with combining the derivatives into a single FM for all $k$- and $z$-bins. Let us assume for illustration purposes the aggressive case, in which  there are 25 $k$-bins for $P_1$ (the linear spectrum in the first bin), and let us forget for now the off-diagonal terms of Eqs. (\ref{eq:offdiagP}) and (\ref{eq:offdiagB}). For each redshift $z_i$, there is a block of $25\times25$ entries for $P_1$, let's call it block $A_i$, a block for the other 15 parameters (a $15\times15$ matrix $C_i$), and a mixed block (plus its transpose), to be called $B_i$, which is a $25\times15$ matrix.  To form the unified FM, blocks $A_i$ are to be simply summed over all the redshifts $z_i$. Blocks $C_i$ have to be aligned along the diagonal, and blocks $B_i$ (and their transpose) have to be aligned along the first 25 rows (or columns, respectively) of the FM, as in the following scheme:
\begin{equation}
    \mathbf{F} = \left(\begin{array}{ccccc}
    \sum_{i}A_{i} & B_{1} & B_{2} & B_{3} & ...\\
    B_{1}^{T} & C_{1} & 0 & 0 & 0\\
    B_{2}^{T} & 0 & C_{2} & 0 & 0\\
    B_{3}^{T} & 0 & 0 & C_{3} & 0\\
    \vdots & 0 & 0 & 0 & \ddots
    \end{array}\right).
    \label{eq:bigmatrix}
\end{equation}
In this example, with $n_b$ $z$-bins, we form a $(25+15n_B)^2$ square FM. The same procedure applies to the bispectrum, padding with zeros the $k$-entries from 0.1 to 0.25 $h/$Mpc.

The off-diagonal terms of Eqs. (\ref{eq:offdiagP}) and (\ref{eq:offdiagB}) introduce  sparse non-zero elements in the ``0''
blocks of (\ref{eq:bigmatrix}). We tested that they introduce very minor changes in the final results (typically improving the constraints by less than 1\%) so we include them only in the $P$ part. For $B$, the derivative wrt $P$ (calculated on the fiducial) is 
\begin{align}
    \frac{\partial B(\mathbf{k}_{1},\mathbf{k}_{2},\mathbf{k}_{3})}{\partial P_1(\mathbf{q})} = \; & 2\big[Z_{1}(\mathbf{k}_{1})Z_{1}(\mathbf{k}_{2})Z_{2}(\mathbf{k}_{1},\mathbf{k}_{2}) \delta_{\mathbf{k}_{1}\mathbf{q}} G^4 P_1(\mathbf{k}_{2})+2\;{\rm cycl.}\big]\\
    & +\big[Z^2_{1}(\mathbf{k}_{1})\delta_{\mathbf{k}_{1}\mathbf{q}}+2\;{\rm cycl.}\big]\frac{1}{n} \,. 
\end{align}
All the other derivatives are straightforward and analytical. 

It is interesting to note that in FreePower  one does not make any assumption on the shape of the power spectrum or bispectrum, not even the presence, or lack thereof, of BAO wiggles in $P(k)$. Nevertheless, the constraining power of the method does depend on the actual shape of $P$ and $B$, and every 
scale dependent feature
introduce additional handles for the AP effect to depend upon. We will assume a standard $\Lambda$CDM fiducial below for all our forecasts.

Before closing this section we comment on how FreePower  is able to recover cosmological information on the Hubble and distance scales, $h$ and $d$, from the AP effect, even after marginalization over $P(k)$ in band-powers. We discussed this issue in detail in Appendix E of \cite{Amendola:2022vte} and we refer the reader there for more details. Considering  $P$, the AP effect distorts the angular dependence only via the $\eta \equiv h \, d$ combination. For a constant growth factor $f$, as considered here, this distortion is degenerate with the amplitude of the monopole of $P$, but cannot be reabsorbed if two or more multipoles (or angular bins) are considered. On the other hand, breaking the degeneracy between $h$ and $d$ requires some information on the scale dependence of $P$.  At the linear level, for a power-law $P$, that is for constant $\partial \log P(k)/\partial \log k $, $h$ and $d$ are exactly degenerate. 
Therefore we need a non power-law $P$ and at least two $k$-bins to break the degeneracy. Including 1-loop corrections, the overall $\Upsilon$ factor is not degenerate with the amplitude of $P(k)$ any more, and this provides a further breaking of the $h-d$ degeneracy.  Adding the bispectrum does not change these general considerations. Considering $B$ alone, breaking the $P-f$ degeneracy requires at least the monopole + quadrupole (or two angular bins). The scale information needed to break the $h-d$ degeneracy comes, once again, from the scale-dependence of the linear $P$, since the non-linear kernels appearing in eq.~(\ref{eq:bisp}) are invariant under a common rescaling of the momenta.

\section{Fiducials and priors}

\setlength\tabcolsep{5.5pt}
\begin{table}[]
    \small
    \centering
    \begin{tabular}{cccccccccccc}
    \hline
    $z$ & $V$  & $n_g\times 10^{-3}$ & $b_1$ & $b_2$ & $c_0$ & $a^{(2)}_\gamma$ & $c^{(2)}_\gamma$  & $d^{(2)}_\gamma$ & $a^{(3)}_{\gamma a}$ & $d^{(3)}_{\gamma a}$ &$\sigma_f$ \\
    &\!\![Gpc$/h]^3\!\!$&  $[h/{\rm Mpc}]^{3}$ & & & $\!\![{\rm Mpc}/h]^2\!$ &&&&&& $\![{\rm Mpc}/h]\!$\\
    \hline
    0.7 & 5.5 & 1.98 & 1.15 & -0.827 & -53 & 1.43 & 1.74 & 0.858 & 0.665 & 0.286 & 3.4 \\
    0.9 & 7.2 & 1.54 & 1.26 & -0.849 & -53 & 1.43 & 1.95 & 0.858 & 0.665 & 0.286 & 3.27 \\
    1.1 & 8.6 & 0.892 & 1.34 & -0.846 & -53 & 1.43 & 2.1 & 0.858 & 0.665 & 0.286 & 3.1 \\
    1.3 & 9.7 & 0.522 & 1.42 & -0.84 & -53 & 1.43 & 2.27 & 0.858 & 0.665 & 0.286 & 2.93 \\
    1.5 & 10.4 & 0.274 & 1.5 & -0.823 & -53 & 1.43 & 2.43 & 0.858 & 0.665 & 0.286 & 2.77 \\
    1.7 & 11.0 & 0.152 & 1.58 & -0.787 & -53 & 1.43 & 2.58 & 0.858 & 0.665 & 0.286 & 2.62 \\
    1.9 & 11.3 & 0.0894 & 1.66 & -0.742 & -53 & 1.43 & 2.75 & 0.858 & 0.665 & 0.286 & 2.47 \\
     \hline
    \end{tabular}
    \caption{Our forecast specifications and fiducials. The volume and density numbers are based on the Euclid survey \cite{Yankelevich:2018uaz}, and  are expected to be similar to those of the DESI survey \cite{DESI:2016fyo}.}
    \label{tab:Euclid}
\end{table}

The fiducial values for $P_1(k_i),\,f,\,H,$ and $D$ are the standard
$\Lambda$CDM ones. We adopt the
following  values: $\Omega_{c0}=0.270$,
$\Omega_{b0}=0.049$, $\Omega_{k0}=0$, $h=0.67$, $n_{s}=0.96$, and
$\sigma_{8}=0.83$. Moreover, we use the approximation $f(z)=\Omega_{m}(z)^{\gamma}$
with $\gamma=0.545$. For the shot-noise parameters $P_{{\rm sn}},B_{\mathrm{sn}(1)},B_{\mathrm{sn}(2)}$
the fiducial is 0.  For $b_1,b_2,c^{(2)}_\gamma$ and $\sigma_f$ we adopted the fiducials in \cite{Yankelevich:2018uaz} (notice  $c^{(2)}_\gamma$ is related to $b_{\mathcal{G}_2}$, see eq.~(\ref{biasvoc})); the fiducials for the PT kernels parameters of eq.~(\ref{beds23}), are the corresponding Einstein-de Sitter values, see eq.~(\ref{EdSvals}).  Finally, for $c_0$ we have adopted the BOSS NGC value at $z=0.61$ of \cite{Ivanov:2019pdj}.  These values are summarized in Table~\ref{tab:Euclid}.

We take infinite uniform priors (i.e.~no prior in the Fisher formalism) for all parameters except for $\log\sigma_{f}$ and the three shot noise parameters, for which we take a Gaussian prior $\mathcal{N}(0,1)$. This amounts to a relative prior of 100\% on $\sigma_{f}$ and on $1+P_{\mathrm{sn}}$ and similarly for the other two shot noise parameters. For $D$ we assume that in each redshift bin a set of SNIa is measured resulting in an effective distance precision of at least 3\%. Since a single type Ia SN yields currently a typical 7\% statistical error, this is a simple requirement of at least 6 events per redshift bin. For $z\le 1$ current constraints from Pantheon+~\cite{Brout:2022vxf} are already at this level. To wit, its binned statistical uncertainties in redshift bins with the same width as ours ($\Delta z = 0.2$) are  \{0.21\%, 0.29\%, 0.49\%, 0.77\%, 3.5\%\} for $z=\{ 0.1,\, 0.3,\, 0.5,\, 0.7,\, 0.9\}$, respectively. LSST on the other hand should achieve even better precision for these bins; for higher $z$'s their proposed survey strategies have much less events~\cite{LSSTScience:2009jmu}. Thus for higher $z$'s, the Roman Space Telescope is the probably the safest bet. It alone is expected to measure distances with statistical errors within $1\%$ in narrower bins for $0.4<z<2.5$~\cite{Rose:2021nzt}, a range exceeding that of either DESI or Euclid data. 
Earlier works have also proposed a high-$z$ Euclid survey, which would measure events up to $z\sim 1.5$. Current surveys are also observing SNIa up to $z\simeq 1.5$~\cite{Yasuda:2019vrq}.

\begin{figure}
    \centering
    \includegraphics[width = .6\textwidth]{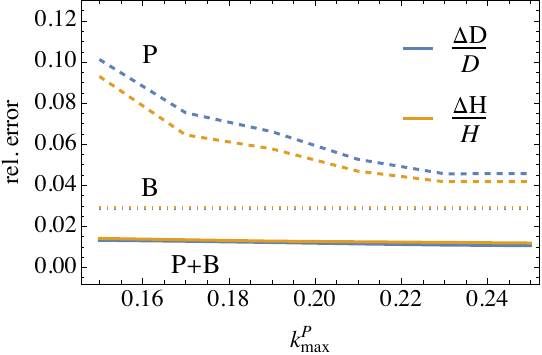}
    \caption{  Trend of the relative errors  on $H$ and $D$  versus $k^P_{{\rm max}}$ for $z=1.1$ when no prior on $D$ is enforced. Dashed lines refer to $P$ alone, dotted lines  to $B$ (with $k^B_{{\rm max}}=0.1h/$Mpc), full lines to $P+B$.  }
    \label{fig:prior-da}
\end{figure}

\begin{figure}
  \centering
  \begin{tikzpicture}
    \node (img1)  
      {\includegraphics[width=0.6\textwidth]{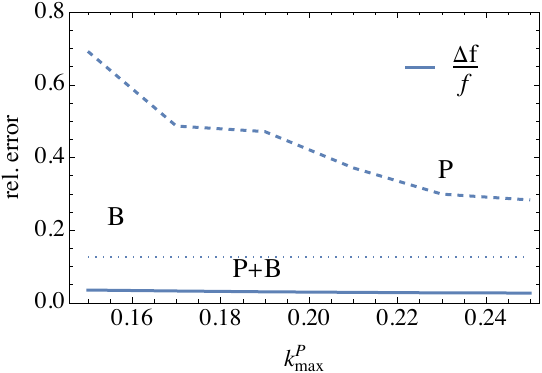}};
    \node[below=of img1, node distance=0cm, yshift=1.5cm, xshift=1.35cm, font=\color{black}] 
      {\tiny };   
  \end{tikzpicture} 
  \caption{  Trend of the relative errors  on $f$  versus $k^P_{{\rm max}}$ for $z=1.1$ when no prior on $D$ is enforced. The dashed line refers to $P$ alone, the dotted line  to $B$ (with $k^B_{{\rm max}}=0.1h/$Mpc), the full line to $P+B$. }
   \label{fig:prior-f}
\end{figure}

If we lift the prior on $D$, we find that the $P$ constraints for $H,D$ remain above 4\% 
as we show in Figure~\ref{fig:prior-da} for the redshift bin $z=1.1$. The final combined result is however prior-independent already at $k_{\rm max} \approx 0.15 h/$Mpc. This remains approximately true for all redshift bins except the farthest one, which is prior-limited even at $k_{\rm max} = 0.25 h/$Mpc. In Figure~\ref{fig:prior-f} we also show the trend of the relative $f$ errors, again when no prior on $D$. In both figures, the strong improvement due to the $P+B$ combination is evident. Figure~\ref{fig:prior-da} and \ref{fig:prior-f} also show that the constraints from the $B$-alone case are better than those in the $P$-alone analysis for all the $k_{\rm max}^{P}$. This is due to the high number of triangles analysed with the anisotropic bispectrum compared to the lower number of $(k, \mu)$ bins of the power spectrum. Moreover, in the one-loop power spectrum $f$ is strongly degenerate with the higher order bootstrap functions $a_\gamma^{(2)}, a_{\gamma a}^{(3)}$ and $d_{\gamma a}^{(3)}$ (which do not enter $B$), leading to a relatively higher error when also these parameters are varied. On the other hand, fixing one or more of these parameters lead to $P$-alone errors on $f$ close to, or better than, the $B$-alone constraints.

\section{Forecasts for the Euclid survey}

We will focus our forecasts for the Euclid space survey, which has been recently launched and will map a sky area of 15000 deg$^{2}$ \cite{laureijs2011euclid}. We remark that  similar constraints should be achieved by the ongoing DESI ground-based survey~\cite{DESI:2016fyo,Vargas-Magana:2018rbb}, which will produce a spectroscopic map covering 14000 deg$^{2}$ of the sky.   The seven redshift bins we employ here and their main properties are in Table \ref{tab:Euclid} (see \cite{2020A&A...642A.191E} and \cite{Yankelevich:2018uaz}). We adopt redshift bins of width $\Delta z=0.2$ equally spaced in the range $[0.6, \, 2.0]$ and assume negligible cross-correlation between them. In our Fisher Matrix calculations, the $\Delta\mu$ step is always fixed to 0.1 and the $\Delta k$ step to 0.01 $h/$Mpc, resulting in up to 55,000 triangle bins and 250 vector bins (plus symmetric configurations). In the first FreePower paper~\cite{Amendola:2019lvy} we adopted $k_{\rm min} = 2\pi / V^{1/3}$, a simple choice which for Euclid translates to $k_{\rm min}$ ranging from $0.0035 \,h/$Mpc, in the first $z$ bin, down to $0.0027\,h/$Mpc in the last. Here instead we take a more conservative value of $k_{\rm min}=0.01 h/$Mpc. An even more conservative choice would be to take $k_{\rm min}$ to correspond to the smallest possible wavelength in a redshift bin so as to completely remove any possible correlations among redshift bins due to large scale modes. For an unmasked observation cone we would have $k_{\rm min} = 2\pi / [D_{\rm C}(z_{\rm bin}+\Delta z/2) - D_{\rm C}(z_{\rm bin}-\Delta z/2)]$, where $D_{\rm C}$ is the line-of-sight comoving distance. For Euclid, this would correspond to $k_{\rm min}$ ranging from $0.016 \,h/$Mpc, in the first $z$ bin, down to $0.030\,h/$Mpc in the last. We have explicitly checked that our results here would change by less than 5\% with this more conservative choice. We note, however, that if one would also add peculiar velocity field measurements, the sensitivity to $k_{\rm min}$ would increase, as discussed in~\cite{Garcia:2020qah}.

We have chosen two combinations of $k_{\rm max}$: a {\it conservative} (pessimistic) one ($k^P_{\rm max}=0.20 \,h/$Mpc for $P$ and $k^B_{\rm max}=0.08 \,h/$Mpc for $B$) and an {\it aggressive} (optimistic) one ($k^P_{\rm max}=0.25 \,h/$Mpc for $P$ and $k^B_{\rm max}=0.1 \,h/$Mpc for $B$).  Similar $k_{\rm max}$ values have been often chosen in forecasts and in real data analysis (e.g.~\cite{Yankelevich:2018uaz,Gualdi:2020ymf,Agarwal:2020lov}), because they are supposed to lie within the regime in which non-linear effects are still not too strong to require even higher-order expansions. It is important to stress that the choice of $k_{\rm max}$ has a crucial impact on the final constraints, even more so when applying our model-independent approach, as illustrated in~\cite{Amendola:2022vte}. The scales at which accuracy loss becomes relevant remains a focus of current research in the field, for instance making use of blind challenges and simulations~\cite{Nishimichi:2020tvu,Brieden:2022ieb}. We nevertheless consider that the two lower $k_{\rm max}$ values, $0.2$ and $0.08\,h/$Mpc are quite conservative (especially at the higher redshifts here considered), while the two upper values, $0.25$ and $0.1\,h/$Mpc, are somehow optimistic.

\setlength\tabcolsep{3.4pt}
\begin{table}[]
\footnotesize
\centering
\begin{tabular}{ccccccccccccccccc}
    \hline\\[-8pt]
    $z$ & $\log f$ & $\log H$ & $\log D$ & $\log b_1$ & $b_2$ & $c_0$ &$a^{(2)}_\gamma$  & $\log c^{(2)}_\gamma$  & $\log d^{(2)}_\gamma$ & $a^{(3)}_{\gamma a}$ & $d^{(3)}_{\gamma a}$ &$\log\sigma_f$ & $P_{\rm sn}$ &  $B_{\rm sn(1)}$ & $B_{\rm sn(2)}$  \\[2pt]
    \hline\\[-8pt]
    \multicolumn{16}{c}{conservative case ($k_{\rm max}^P=0.2 \,h/$Mpc, $k_{\rm max}^B=0.08 \,h/$Mpc)}\\[2pt]
    \hline    
  0.7 & 0.039 & 0.018 & 0.013 & 0.032 & 0.18 & 5.9 & 7.3 & 0.13 & 0.19 & 19. & 8.2 & 0.19 & 0.74 & 0.72 & 1. \\
 0.9 & 0.038 & 0.016 & 0.013 & 0.031 & 0.19 & 6. & 8.2 & 0.12 & 0.2 & 22. & 9.2 & 0.18 & 0.6 & 0.58 & 1. \\
 1.1 & 0.038 & 0.016 & 0.012 & 0.032 & 0.21 & 6.7 & 9.6 & 0.13 & 0.23 & 26. & 11. & 0.18 & 0.36 & 0.38 & 1. \\
 1.3 & 0.039 & 0.016 & 0.012 & 0.033 & 0.25 & 7.8 & 11. & 0.14 & 0.28 & 31. & 13. & 0.19 & 0.21 & 0.26 & 0.96 \\
 1.5 & 0.045 & 0.017 & 0.013 & 0.037 & 0.34 & 9.9 & 14. & 0.18 & 0.39 & 38. & 15. & 0.21 & 0.12 & 0.24 & 0.77 \\
 1.7 & 0.057 & 0.021 & 0.015 & 0.043 & 0.51 & 14. & 18. & 0.25 & 0.59 & 50. & 19. & 0.26 & 0.073 & 0.25 & 0.46 \\
 1.9 & 0.072 & 0.026 & 0.017 & 0.051 & 0.83 & 20. & 24. & 0.37 & 0.96 & 67. & 26. & 0.35 & 0.052 & 0.24 & 0.27 \\   
    \hline\\[-8pt]
    \multicolumn{16}{c}{aggressive case ($k_{\rm max}^P=0.25 \,h/$Mpc, $k_{\rm max}^B = 0.10 \,h/$Mpc)}\\[2pt]
 \hline
  0.7 & 0.026 & 0.012 & 0.0088 & 0.021 & 0.1 & 3.7 & 3.7 & 0.082 & 0.11 & 9.6 & 4.1 & 0.072 & 0.56 & 0.35 & 1. \\
 0.9 & 0.025 & 0.011 & 0.009 & 0.021 & 0.11 & 4.2 & 4.1 & 0.078 & 0.12 & 11. & 4.5 & 0.074 & 0.43 & 0.26 & 1. \\
 1.1 & 0.025 & 0.011 & 0.0092 & 0.022 & 0.13 & 5. & 4.6 & 0.083 & 0.14 & 12. & 5.1 & 0.081 & 0.25 & 0.18 & 0.97 \\
 1.3 & 0.026 & 0.012 & 0.0095 & 0.024 & 0.16 & 5.9 & 5.4 & 0.097 & 0.17 & 14. & 5.9 & 0.09 & 0.15 & 0.17 & 0.8 \\
 1.5 & 0.031 & 0.014 & 0.01 & 0.027 & 0.22 & 7.7 & 6.8 & 0.13 & 0.25 & 18. & 7.4 & 0.11 & 0.083 & 0.16 & 0.43 \\
 1.7 & 0.039 & 0.018 & 0.012 & 0.031 & 0.34 & 11. & 9.2 & 0.18 & 0.4 & 25. & 9.9 & 0.14 & 0.055 & 0.15 & 0.21 \\
 1.9 & 0.05 & 0.023 & 0.015 & 0.037 & 0.56 & 16. & 13. & 0.26 & 0.68 & 36. & 14. & 0.2 & 0.041 & 0.15 & 0.12 \\

   \hline
    \end{tabular}
\caption{ Euclid marginalized parameter forecasts for all redshift bins. }
\label{tab:Euclid-results}
\end{table}

Table~\ref{tab:Euclid-results} summarizes our final, marginalized, results for all the $z$-dependent parameters   in both conservative and aggressive cases. It shows only the combined $P+B$ forecasts, i.e. where the bispectrum is included. We make comparisons to the case without bispectrum below. It is interesting to note that this method yields precise measurements of the growth rate $f$ independently of $\sigma_8$, achieving 2--3\% (4--5\%) precision for the aggressive (conservative) case, considering the lowest five redshift bins (all the constraints mentioned in the following also refer to these bins). The results on the Hubble function and  distance are also very precise, typically at the 1\% level and sometimes  better. Regarding the chosen priors, in both conservative and aggressive cases the informative priors on the shot-noise parameter $B_{\rm sn(2)}$ is dominant at the lower redshift bins. The priors on the other shot noise parameters are on the other hand not driving any other results, which is reassuring. The 3\% priors on distance are also shown not to be dominant, and constraints are always at least twice as precise as it. Ref.~\cite{Philcox:2022frc} has already shown that including the bispectrum improves significantly the precision on higher order biases, in particular on $b_2$ and $b_{\mathcal{G}_2}$, the latter being related, in our notation, to $a_\gamma^{(2)}$ and $c_\gamma^{(2)}$, see eq.~\eqref{biasvoc}. Here we show that the bispectrum greatly improves the constraints for all the non-linear parameters, both  the tracer-independent ones, in particular $d_\gamma^{(2)}$, and the bias parameters $b_2$ and $c_\gamma^{(2)}$. For the second order bias $b_2$ we reach an absolute accuracy of 0.1--0.2 (0.2--0.3)  while for $c_\gamma^{(2)}$ we find 8--13\% (13--18\%) errorbars in the aggressive (conservative) case. These are in line with results of~\cite{Philcox:2022frc}, which performed an MCMC analysis on large volume numerical simulations using the 1-loop model for the bispectrum. We report here the first results on the bootstrap parameter $d_\gamma^{(2)}$, for which we reach an accuracy of 12--25\% (20--40\%). These constraints degrade at higher redshifts, where the nonlinear effects become less and less relevant.

\begin{figure}
    \centering
    \includegraphics[trim={0cm 0cm 0cm 0cm}, clip, width = 0.7\textwidth]{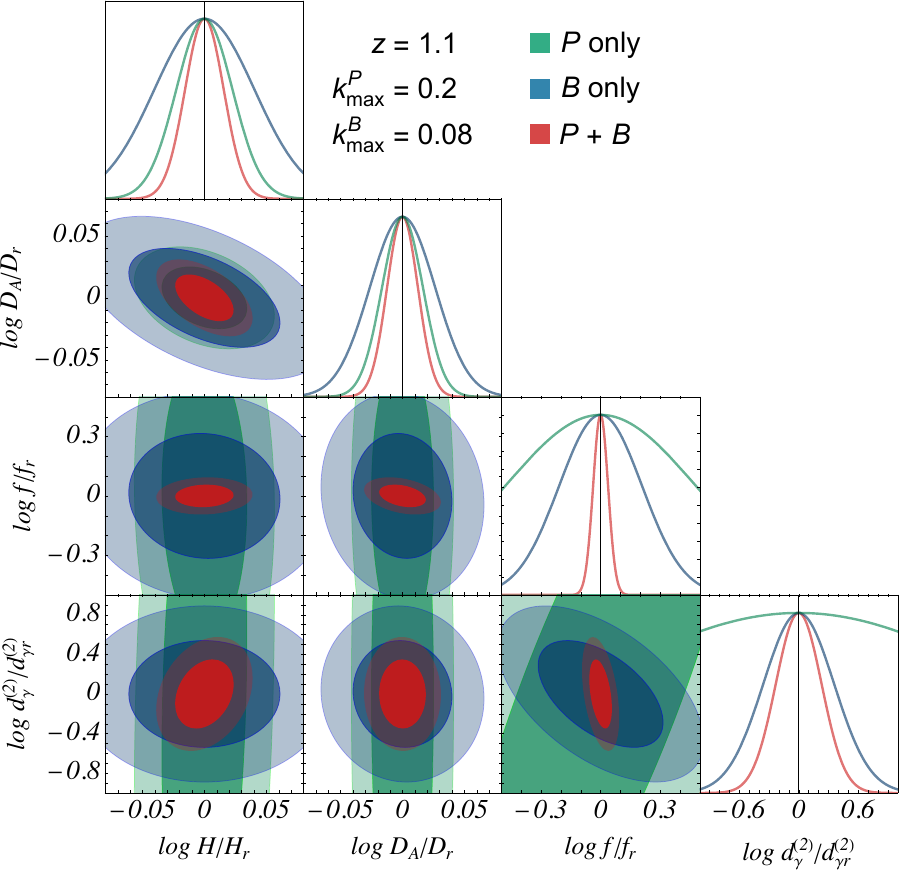}
    \includegraphics[trim={0cm 0cm 0cm 0cm}, clip, width = 0.7\textwidth]{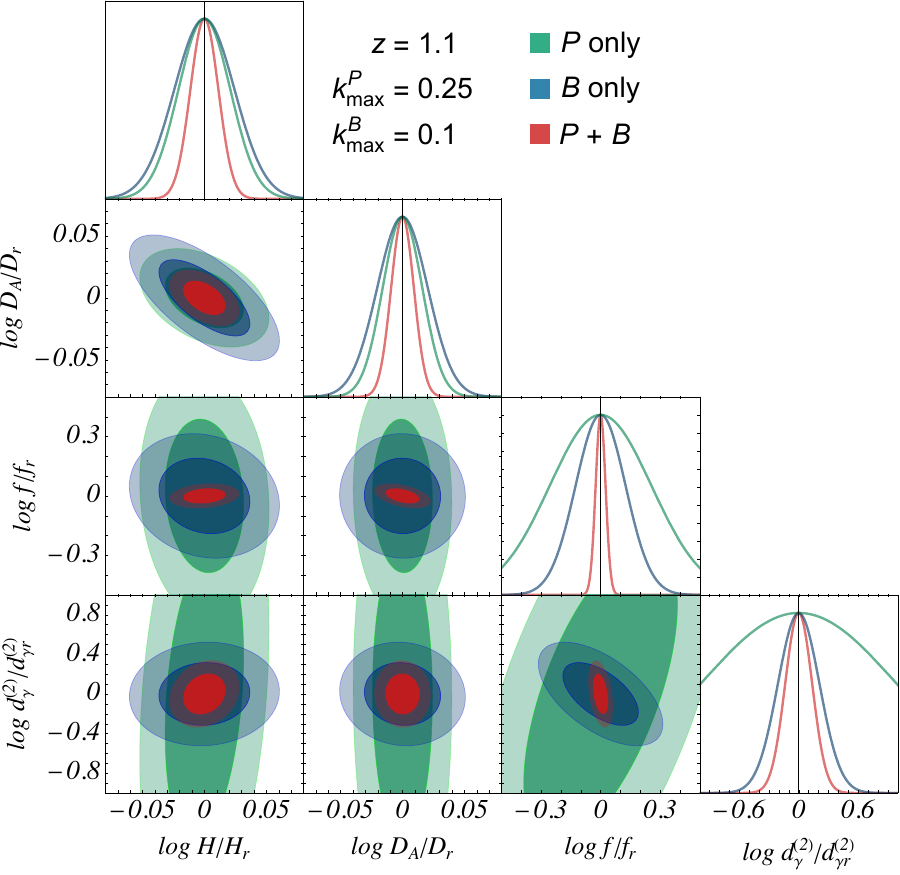}    
    \caption{ Results for the Fisher analysis of a Euclid-like survey at $z=1.1$ for the tracer-independent parameters.  \emph{Top:} conservative case. \emph{Bottom:} aggressive case.}    
    \label{fig:Euclid-grid-tracer-indep}
\end{figure}

\begin{figure}
    \centering
  \includegraphics[trim={0cm .2cm 0cm 1.2cm}, clip, width = 0.73\textwidth]{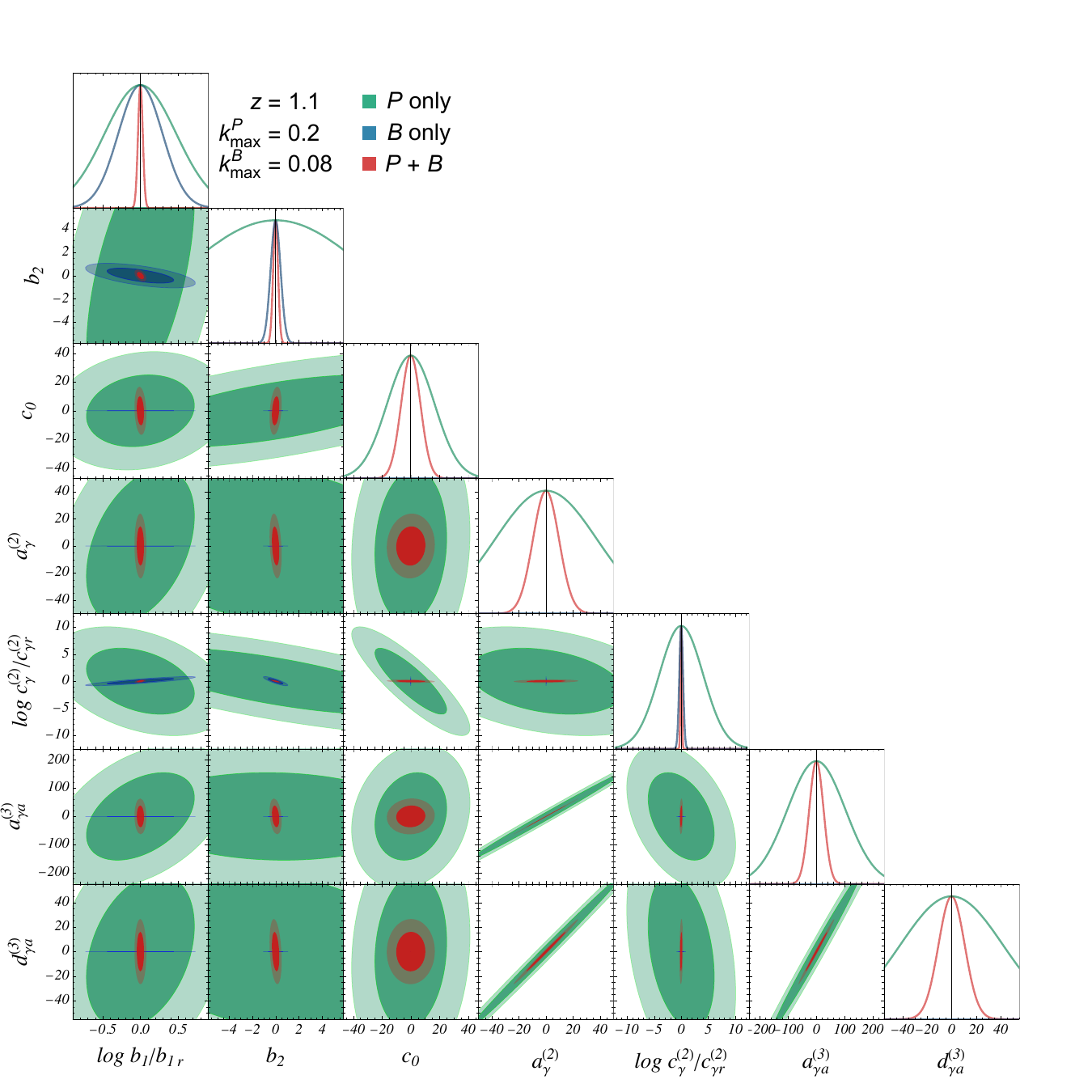}
    \includegraphics[trim={0cm .1cm 0cm 1.2cm}, clip, width = 0.73\textwidth]{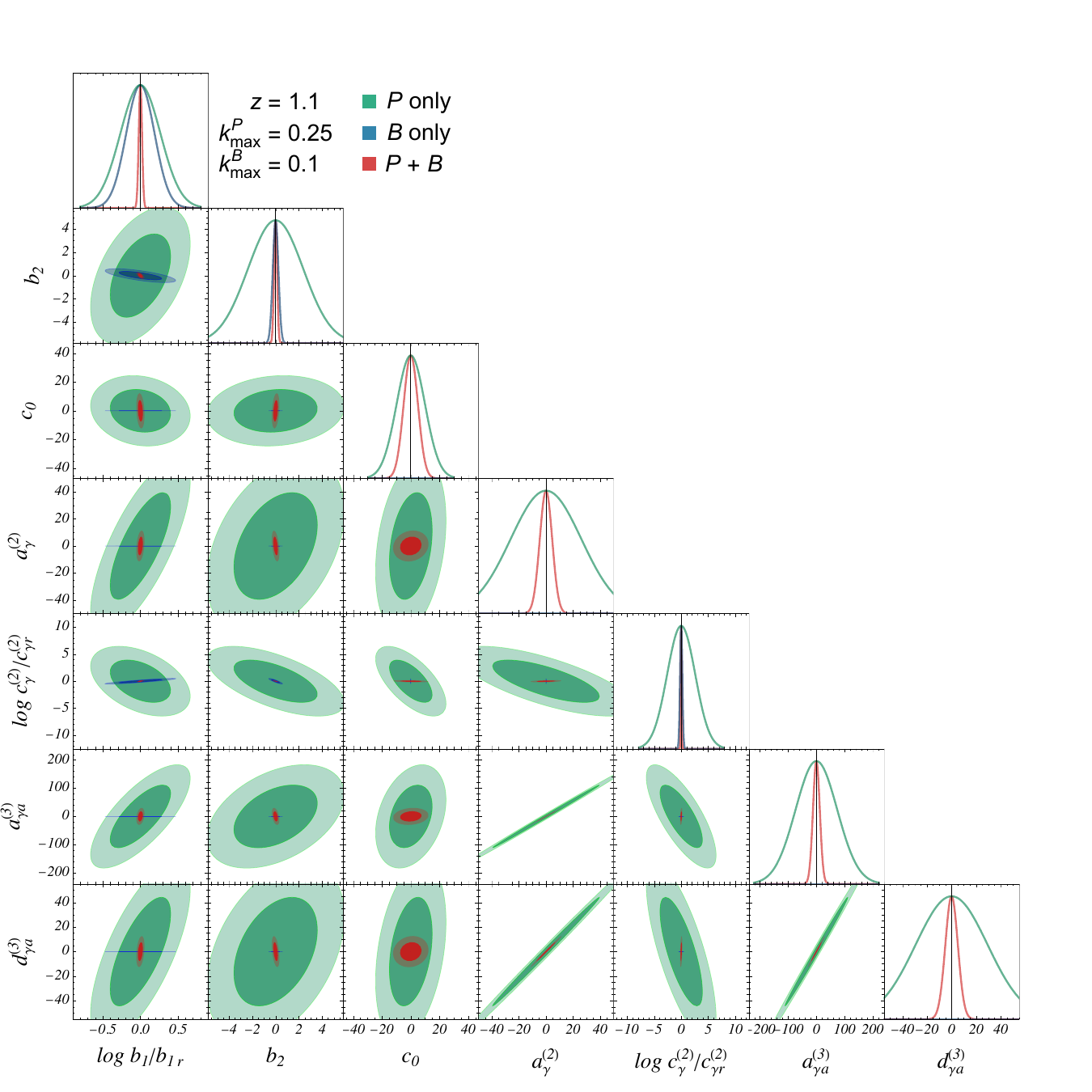}    
       \caption{ Same as Figure~\ref{fig:Euclid-grid-tracer-indep}, but for the tracer-dependent parameters. Note that $B$ does not depend on $c_0, a^{(2)}_{\gamma},a^{(3)}_{\gamma a}, d^{(3)}_{\gamma a}$, which have been therefore fixed to their fiducial when drawing $B$ contour plots.    }
    \label{fig:Euclid-grid-tracer-dep}
\end{figure}

In Figures~\ref{fig:Euclid-grid-tracer-indep} and~\ref{fig:Euclid-grid-tracer-dep} we show the 68.3 and 95.4\% confidence regions at $z=1.1$ for a selection of parameters, namely the tracer-independent ones ($H,D,f,d^{(2)}_\gamma$), and the tracer-dependent ones ($b_1,b_2,a^{(2)}_\gamma,c^{(2)}_\gamma,a^{(3)}_{\gamma a},d^{(3)}_{\gamma a}$). In these as in all other plots, the errors are marginalized over all the other parameters. Notice that the $P$+$B$ FM includes a single set of priors, not two sets, and therefore is not simply the sum of the $P$ and $B$ FMs. These figures illustrate how much the addition of the bispectrum can increase the precision of some parameters. It also shows that some of the PT kernels parameter and the higher order biases are highly correlated, and that the bispectrum breaks some of these degeneracies. The improvement on the second order perturbative  parameters, $b_2$, $c_\gamma^{(2)}$ and $d_\gamma^{(2)}$, which contribute to the tree-level bispectrum, is dramatic. In particular, reducing the errors on the cosmology dependent function $d_\gamma^{(2)}$ could lead to constraints on deviations from the standard cosmological model from perturbations at second order, as we comment after Figure~\ref{fig:Euclid-fz}. The parameters appearing at third order, namely, $a_\gamma^{(2)}$, $a_{\gamma a}^{(3)}$ and $d_{\gamma a}^{(3)}$, are strongly degenerate. Adding the bispectrum improves indirectly their errorbars by breaking some of degeneracies, but not enough to constrain them, as these parameters do not enter the tree level expression for the bispectrum. This is usually also the case for standard model-dependent analysis in which these parameters are fixed or analytically marginalized. In appendix \ref{app:full_tri} we show for completeness the contour plots for all the parameters.

\begin{figure}
    \centering
     \includegraphics[width = 0.47\textwidth]{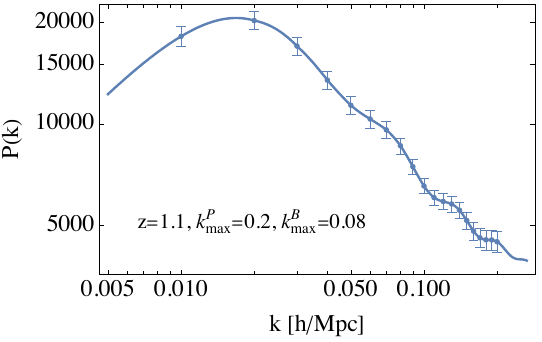}
    \includegraphics[width = 0.47\textwidth]{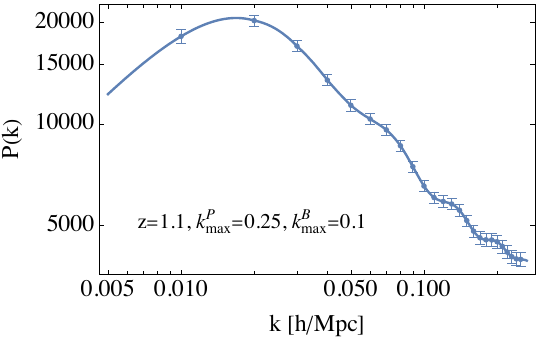}
    \caption{  Forecasts for $P(k)$ (with $\mu=0.5$) for an Euclid-like survey using $P+B$, for the arbitrary choice of $z=1.1$ as our pivot redshift (see text). Note that the error bars have important correlations. \emph{Left:} conservative case; \emph{Right:} aggressive case. } 
    \label{fig:Euclid-pk}
\end{figure}

In Figure~\ref{fig:Euclid-pk} we illustrate the forecast uncertainties on the power spectrum $k$-bins for both aggressive and conservative cases.  As discussed above, FreePower  does not assume a specific shape for $P(k)$. Nevertheless it is capable of reconstructing the power spectrum in all $k$-bins including the wiggles with good precision: we find errors in the range 3--5\% for the aggressive case, and 5--7\% for the conservative one.  We remark however that these bins have important correlation in the error bars; we quantify this below. Note that in FreePower  we assume a single set of linear $P(k)$ at a given ``pivot'' redshift and relates it to the linear spectrum at other redshifts through the growth-rate $f$. We choose in this plot the intermediate redshift bin $z=1.1$ as our pivot, but this choice only affects this particular plot, and our forecasts are invariant over this choice. It is interesting to note that the inclusion of the bispectrum data indirectly improves the precision on the $P(k)$ by a large amount due to the breaking of important degeneracies. The average precision gains among all the $k$-bins is an impressive factor of 7 (11) in for the aggressive and the (conservative) case. More importantly, these breaking of degeneracies also lead to less correlations.

\begin{figure}
    \centering 
    \includegraphics[width = 0.47\textwidth]{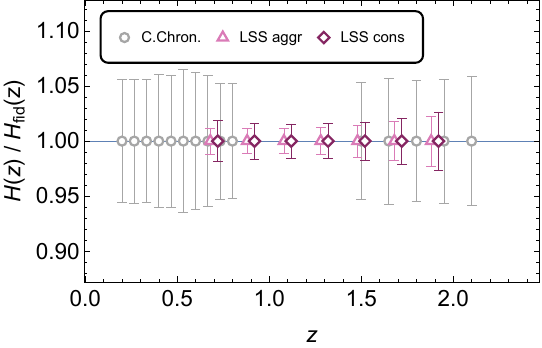}\quad
    \includegraphics[width = 0.47\textwidth]{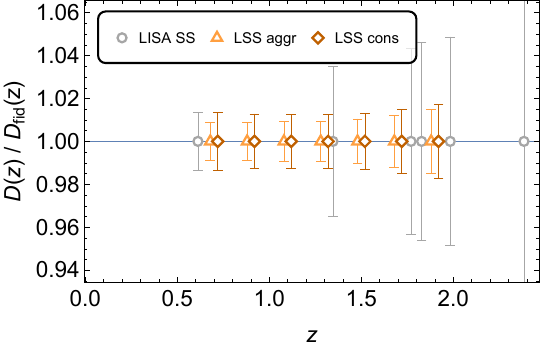}
    \caption{  Forecast comparison for the expansion rate and distances of our Euclid $P+B$ (both aggressive and conservative cases) with those of different probes. We add slight displacements in $z$ for clarity. \emph{Left:} $H$ forecasts comparing with cosmic chronometers~\cite{Moresco:2022phi}. \emph{Right:} $D$ forecasts comparing with  LISA standard sirens~\cite{Speri:2020hwc}. Correlations among LSS data pairs are around 0.12 (0.34) for $H$ ($D$). 
    }
    \label{fig:Euclid-HDz}
\end{figure}

In order to contextualise our background forecasts with other proposed probes, we combine in Figure~\ref{fig:Euclid-HDz} our predictions with a couple of other forecasts in the literature. For the Hubble function, there are not many direct (cosmology-independent) probes. We focus on the cosmic chronometers, the use of which has become more widespread in recent years. Recently, forecasts for future  data were performed in~\cite{Moresco:2022phi}. Cosmic chronometers have many possible source of systematics related to astrophysical model assumptions, but these  are expected to be unrelated to the background cosmology. The redshift drift is also an alternative, but lies farther ahead in the future~\cite{Moresco:2022phi}. As can be seen, our forecasts compare favourably with cosmic chronometers even in our conservative case. We also note that in this and all figures that follow the errors in different redshift bins are only lightly correlated, as we will show below.

For distances, the classic cosmology probe are type Ia supernovae. By themselves, however, they are only capable of measuring relative distances (they require external calibration in order to measure absolute distances), and moreover it is currently a great challenge for systematic uncertainties to keep up with the expected decrease in the statistical ones ~\cite{Hounsell:2017ejq} since they are already currently at comparable levels~\cite{Brout:2022vxf}. Bright standard sirens, on the other hand, measure absolute distances and the LISA space detector is expected to detect them at high redshifts. We thus include a forecast of fifteen events made in Ref.~\cite{Speri:2020hwc} in the right panel of Figure~\ref{fig:Euclid-HDz}. FreePower's forecasts  again compare favourably but, on the other hand, we note that LISA is in principle capable  of detecting bright sirens to much higher redshifts, up to around $6$ (not shown in our plot).

\begin{figure}
    \centering 
    \includegraphics[trim={5.1cm 0cm 5.2cm 0cm}, clip, width = 0.95\textwidth]{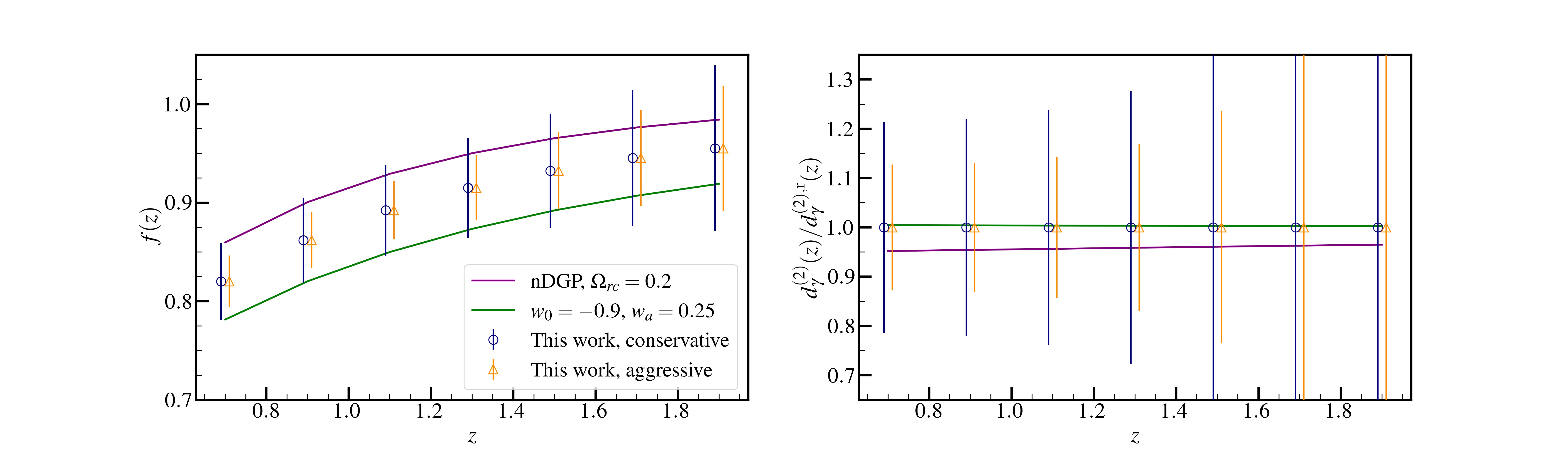}
   \caption{Similar to Figure~\ref{fig:Euclid-HDz} but for the 
    functions $f(z)$ and $d_\gamma(z)$ at different redshifts. Here we plot two possible alternatives to the standard $\Lambda$CDM cosmology: the normal branch of the DGP model~\cite{Dvali:2000hr}, which was constrained recently using the BOSS data~\cite{Piga:2022mge} and  an evolving dark energy with equation of state $w(z) = w_0 + w_a z/(1+z)$. Correlations among the error bars are small. 
    }
    \label{fig:Euclid-fz}
\end{figure}

Figure~\ref{fig:Euclid-fz} shows the results for the marginalized 1-$\sigma$ errorbars for the growth function $f(z)$ and the bootstrap function $d_\gamma^{(2)}(z)$. In these plots we also show, for comparison, some models beyond $\Lambda$CDM that are still phenomenologically viable: the normal branch of the DGP model~\cite{Dvali:2000hr}, which was constrained against the BOSS data recently~\cite{Piga:2022mge}, and an evolving dark energy with equation of state $w(z) = w_0 + w_a z/(1+z)$. Our analysis shows impressive results for the growth function: the $P$+$B$ analysis has the full potential to distinguish among different cosmological models. Exotic $w_0-w_a$ models, like the ones that are found to fit quasars observations~\cite{Risaliti:2018reu,Bargiacchi:2021hdp}, will potentially be constrained by the next generation of LSS surveys. The nDGP parameter $\Omega_{\rm rc}$ could potentially be constrained at the percent accuracy without assuming any model and any tight prior on the parameters of the primordial power spectrum as in the analysis performed in~\cite{Piga:2022mge}.  The modifications on the nonlinear parameter $d_\gamma^{(2)}$ are typically smaller than the errorbars of our fully model-independent analysis. However, these effects could be captured in a more constrained analysis aimed at detecting late-time modifications of $\Lambda$CDM, in particular involving extra scalar degrees of freedom and screening mechanisms, which would typically affect the growth and the nonlinear kernels while leaving the power spectrum shape unchanged.

\begin{figure}
    \centering 
    \includegraphics[width = 0.54\textwidth]{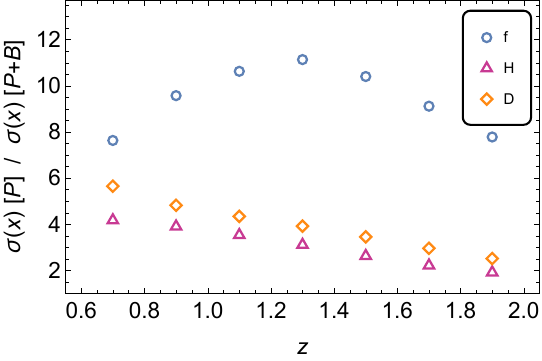}      
    \caption{Comparisons of constraints in $f$, $H$ and $D$ (all generically labeled as ``$x$'' in the label) from $P$ alone and $P$+$B$ for the aggressive case without $D$ priors. As can be seen, the bispectrum greatly improves the precision in $f$ in all $z$ bins, while the gains in $H$ and $D$ are more modest.
    }
    \label{fig:fHD-P-PB}
\end{figure}

Figure~\ref{fig:fHD-P-PB} illustrates more in detail how much precision the bispectrum adds to the classic cosmological parameters $f$, $H$ and $D$ in all redshift bins considered. In this plot and the next we remove the 3\% prior on $D$ in order to have a clearer picture on how much information the bispectrum is contributing.
For the background parameters the gains are 
significant, especially at lower redshifts, and on average across all $z$s a factor between 3 and 4. For the growth rate $f$ the increase is even larger, to wit on average a factor higher than 9, peaking around $z = 1.3$.

\begin{figure}
    \centering 
    \qquad\qquad\includegraphics[width = 0.68\textwidth]{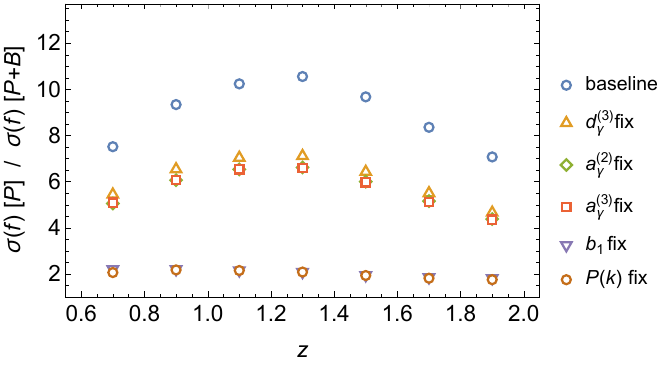}   
    \caption{
    Similar to Figure~\ref{fig:fHD-P-PB} for $f$ while fixing different variables to their fiducial value. As can be seen, most of the relative gains in $P$+$B$ comes from the extra constraints on either $b_1$ and $P(k)$, followed by the constraints on the trio $a_\gamma^{(2)}$, $a_\gamma^{(3)}$ and $d_\gamma^{(3)}$, which are very degenerate among themselves.
    }
    \label{fig:f-P-vs-PB}
\end{figure}

This precision increase is in general due to the fact that adding the bispectrum results in breaking important parameter degeneracies, a point to which we will return below. But it is interesting to explore deeper why the increase is so much larger in $f$ than in $H$ or $D$.  So in Figure~\ref{fig:f-P-vs-PB} we explore what happens if we fix some of the parameters by hand (here again we remove the 3\% $D$ prior). We see that most of the relative gains in $P$+$B$ comes from the extra constraints on  on either $b_1$ and $P(k)$, followed by the constraints on the trio $a_\gamma^{(2)}$, $a_\gamma^{(3)}$ and $d_\gamma^{(3)}$, which are very degenerate among themselves. The extra constraints on $b_2$ are also important but to a lesser degree. We find that, when $B(\mathbf{k}_1, \mathbf{k}_2, \mathbf{k}_3)$ is not included, $f$ has significant correlations with $b_1$ (typically around 0.97) and the trio $a_\gamma^{(2)}$, $a_\gamma^{(3)}$ and $d_\gamma^{(3)}$ (between 0.5 and 0.8 depending on redshift). It is the suppression of these four correlations, together with the suppression of the correlations between $f$ and some $k$-bins of $P$, what drives the gains in $f$ in $P$+$B$. Concerning $b_1$ and $P(k)$, both also become much better measured, respectively by factors of 11 (averaged over all $z$s) and 7 (averaged over all $k$s).

\begin{figure}
    \centering 
    \includegraphics[width = 0.79\textwidth]{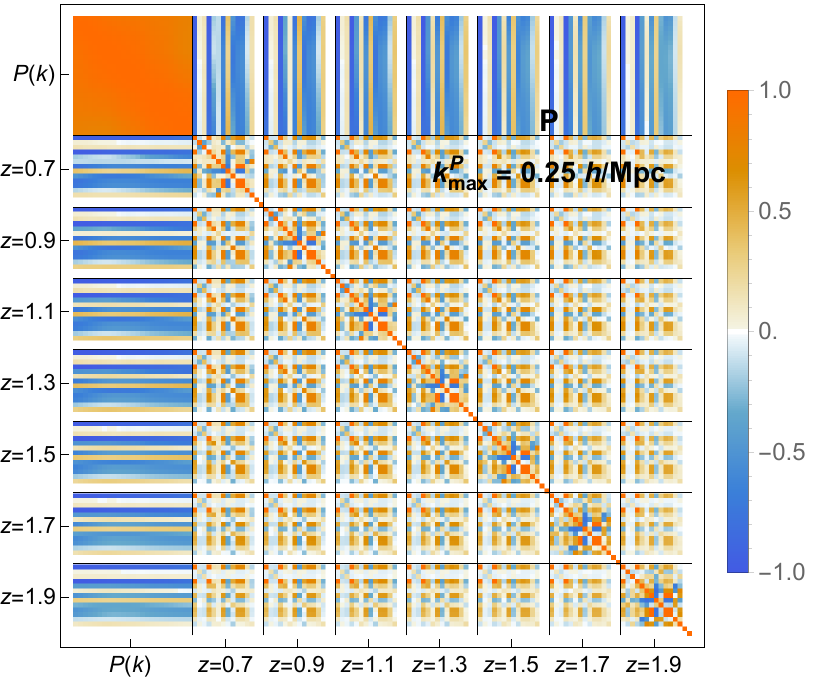}
    \includegraphics[width = 0.79\textwidth]{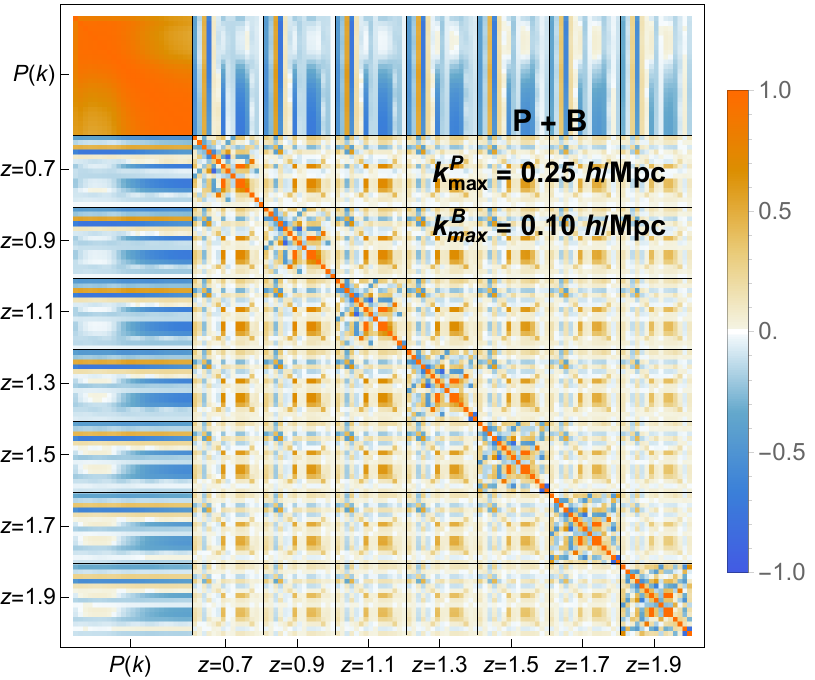} 
    \caption{Full correlation matrix of all our 130 variables for the aggressive case using only $P$ (top) and $P$+$B$  (bottom). Note that correlations are typically smaller at higher $z$ and for $P$+$B$. In each $z$ bin the parameter order is this: $\{\log f,\log H,\log D,\log b_{1},b_{2}, c_{0}, \,a_{\gamma}^{(2)}, \,\log c_\gamma^{(2)}, \,\log d_{\gamma}^{(2)},\, a_{\gamma a}^{(3)}, \,d_{\gamma a}^{(3)}, \log\sigma_{f}, P_{\mathrm{sn}},B_{\mathrm{sn}(1)},B_{\mathrm{sn}(2)}\}$.
    }
    \label{fig:corr-mat}
\end{figure}

Figure~\ref{fig:corr-mat} shows the full correlation matrix of all our 130 variables for the aggressive case using only $P$ and $P+B$. In general one notices larger correlations among the variables in lower redshift bins than in higher ones.
More importantly, the inclusion of the bispectrum greatly reduces correlations in some key variables. In particular, degeneracies involving $f$ and $b_1$, $b_2$, and the trio $a_\gamma^{(2)}$, $a_\gamma^{(3)}$ and $d_\gamma^{(3)}$ are substantially diminished in the $P+B$ case. Correlations among the 15 $z$-dependent parameters in a given $z$-bin are moderate, to wit around 0.23 average absolute correlation for either cases $P$ or $P+B$.\footnote{Using the absolute correlation coefficient value takes into account the fact that both negative and positive correlations increase degeneracies among parameters.} Correlations of $f(z)$ measurements between different redshift bins using $P$ alone are very high, to wit around 0.97. Adding the bispectrum, it collapses to an average of 0.29. Correlations of both background parameters $H$ and $D$  among different redshift bins are mild even with $P$ alone, and we get the following average absolute correlation for $P$ ($P+B$): for $H$, 0.24 (0.12), and for $D$, 0.30 (0.34). This is the reason for our previous statement that the error bars in Figure~\ref{fig:Euclid-HDz} can be assumed to be approximately uncorrelated.

\section{Conclusions}

In this paper we presented a novel methodology, dubbed FreePower, to exploit the increasing amount of information in large scale surveys without restricting to a particular model. We modelled the linear power spectrum by a number of waveband parameters, rather than in terms of cosmological parameters, and left free to vary 15 more parameters per redshift bin related to cosmology, to linear and non-linear biases,  to generalized kernels, and to shot-noise. We have shown that by including the one-loop power spectrum and the tree-level bispectrum it is possible to typically reach, with either the Euclid or DESI surveys, in seven redshift bins from 0.6 to 2.0, a simultaneous $1-3\%$ precision 
in the Hubble function, the distance, the growth rate, and the linear bias.

The potential of the bispectrum in cosmological analyses has been investigated in several papers \cite{Yankelevich:2018uaz, Gualdi:2020ymf, Agarwal:2020lov, Pardede:2022udo}. The general conclusion is that, especially when prior information from cosmic microwave background is included in the analyses, the information added by the bispectrum is mainly confined to a better determination of the bias parameters, while the improvement on the cosmological ones is marginal. Our model-independent analysis, on the other hand, has highlighted that the inclusion of the bispectrum improves the constraints on the linear growth by a factor larger than three (see Figures~\ref{fig:prior-f} and~\ref{fig:fHD-P-PB}).  Moreover, even including extra degrees of freedom in the form of general expressions for the PT kernels we can confirm that the bispectrum improves substantially the determination of tracer-dependent bias parameters, in particular $c_\gamma^{(2)}$ (which is related to $b_{{\cal G}_2}$ in a fixed cosmology). At the same time, non-trivial information on the cosmology, encoded in the tracer-independent parameter $d_\gamma^{(2)}$, can be recovered at the $10-25\,$\% level. It is the first time, to our knowledge, that the possibility of extracting cosmological information from the nonlinear sector of cosmological perturbations at this level is highlighted.

The analysis put forward in this paper can of course be replicated also for testing specific models, in particular $\Lambda$CDM. Clearly, the more restricted is the number of free parameters, the stronger the constraints will be. 

Our approach still has several limitations that can be overcome in the future. For instance, we do not include terms due to the window function or the trispectrum to either the power spectrum  and the bispectrum.
Moreover, we assume  a $k$-independent growth rate. The Fisher matrix approximation might also prove not to be entirely adequate, and it would be important in the future to cross-check it either using the higher-order DALI method~\cite{Sellentin:2014zta} or performing a full MCMC. While overcoming these limitations  may degrade the constraints, there are also several improvements that promise to lead to stronger bounds. For instance, the inclusion of multi-tracers is expected to lead to better constraints on the tracer-independent parameters~\cite{Abramo:2013awa}. Moreover, higher $k_{\rm max}$ values could be employed by adding higher-order kernels.  On a different note, we remark that the estimation of $H(z)$ and $D(z)$ can be employed to measure the spatial curvature regardless of the cosmological model. Some of these topics will be investigated in future work.

\section*{Data availability}

The Mathematica code to produce the calculations of this paper is publicly available on \url{https://github.com/itpamendola/FreePower}.

\section*{Acknowledgements}

We thank Michele Moresco for providing data for Figure~\ref{fig:Euclid-HDz}. LA thanks Hector Gil-Marin and Victoria Yankelevich for useful discussions, and Tjark Harder for discussions on the Fisher code and for detecting a bug in the original version.
LA acknowledges support from DFG project  456622116. MM acknowledges support by the Israel Science Foundation (ISF) grant No. 2562/20. MP acknowledges support by the MIUR Progetti di Ricerca di Rilevante Interesse Nazionale (PRIN) Bando 2022 - grant 20228RMX4A. MQ is supported by the Brazilian research agencies FAPERJ, CNPq (Conselho Nacional de Desenvolvimento Científico e Tecnológico) and CAPES. This study was financed in part by the Coordenação de Aperfeiçoamento de Pessoal de Nível Superior - Brasil (CAPES) - Finance Code 001. We acknowledge support from the CAPES-DAAD bilateral project  ``Data Analysis and Model Testing in the Era of Precision Cosmology''. Several integrals have been performed using the CUBA routines \cite{Hahn_2005} by T. Hahn (http://feynarts.de/cuba). The fiducial $\Lambda$CDM linear power spectrum has been obtained with the CAMB code \cite{Lewis:1999bs}. Some of the triangle plots were performed using the code GetDist~\cite{Lewis:2019xzd}.

\appendix

\section{Kernels beyond Einstein-de Sitter}
\label{app:beds}

To go beyond the usual Einstein-de Sitter approximation, we define the functions 
\begin{align}
\alpha_s(\bq_1,\bq_2) & =1+\frac{\bq_{1}\cdot \bq_{2}}{2}\left(\frac{1}{q_{1}^{2}}+\frac{1}{q_{2}^{2}}\right)\\
\beta(\bq_{1},\bq_{2}) & =\frac{|\bq_{1}+\bq_{2}|^{2}\bq_{1}\cdot \bq_{2}}{2q_{1}^{2}q_{2}^{2}}\\
\gamma(\bq_{1},\bq_{2}) & =1-\frac{(\bq_{1}\cdot \bq_{2})^{2}}{q_{1}^{2}q_{2}^{2}}=\alpha_{s}(\bq_{1},\bq_{2})-\beta(\bq_{1},\bq_{2})\\
\alpha_{a}(\bq_{1},\bq_{2}) & =\frac{\bq_{1}\cdot \bq_{2}}{q_{1}^{2}}-\frac{\bq_{2}\cdot \bq_{1}}{q_{2}^{2}}
\end{align}
We define now  beyond-EdS kernels for matter \cite{DAmico:2021rdb},
\begin{align}
& 2\,F_{2}(\bq_1,\bq_2; z)   =2\, \beta(\bq_{1},\bq_{2})+ a_{\gamma}^{(2)}(z)\,\gamma(\bq_{1},\bq_{2})\,,\\
& 3!\, F_{3}(\bq_1,\bq_2,\bq_3; z)  =2\, O_{\beta\beta}(\bq_{1},\bq_{2},\bq_{3})
+\bigg[a_{\gamma a}^{(3)}(z)\nonumber \\
& \quad -2\left(a_{\gamma b}^{(3)}(z) - 1 \right)\bigg]O_{\beta\gamma}(\bq_{1},\bq_{2},\bq_{3}) +\left(\frac{1}{4} a_{\gamma a}^{(3)}(z)-\frac{1}{2}a_{\gamma b}^{(3)}(z)\right)O_{\gamma\alpha_{a}}(\bq_{1},\bq_{2},\bq_{3})\nonumber\\
 & \quad +\left(a_{\gamma b}^{(3)}(z)-\frac{1}{2}a_{\gamma a}^{(3)}(z)+2(a_{\gamma}^{(2)}(z) - 1)\right) O_{\gamma\beta}(\bq_{1},\bq_{2},\bq_{3}) \nonumber\\
 & \quad  +\left(\frac{1}{2}a_{\gamma a}^{(3)}(z)+a_{\gamma b}^{(3)}(z)\right)O_{\gamma\gamma}(\bq_{1},\bq_{2},\bq_{3})+ {\rm 3\;cyclic}\,   \label{eq:F3},
\end{align}
and for the velocity divergence,
\begin{align}
& 2\,G_{2}(\bq_1,\bq_2; z)   =2\, \beta(\bq_{1},\bq_{2})+ d_{\gamma}^{(2)}(z)\,\gamma(\bq_{1},\bq_{2})\,,\\
& 3!\, G_{3}(\bq_1,\bq_2,\bq_3; z)  =2\, O_{\beta\beta}(\bq_{1},\bq_{2},\bq_{3})
+\bigg[2d_{\gamma}^{(2)}(z)+d_{\gamma a}^{(3)}(z)\nonumber \\&\quad -2\left(d_{\gamma b}^{(3)}(z)+a_{\gamma}^{(2)}(z) - 1\right)\bigg] O_{\beta\gamma}(\bq_{1},\bq_{2},\bq_{3}) +\left(\frac{1}{4}d_{\gamma a}^{(3)}(z)-\frac{1}{2}d_{\gamma b}^{(3)}(z)\right)O_{\gamma\alpha_{a}}(\bq_{1},\bq_{2},\bq_{3})\nonumber\\
&\quad +\left(d_{\gamma b}^{(3)}(z)-\frac{1}{2}d_{\gamma a}^{(3)}(z)+2\left(a_{\gamma}^{(2)}(z) - 1\right)\right) O_{\gamma\beta}(\bq_{1},\bq_{2},\bq_{3})\nonumber\\
&\quad +\left(\frac{1}{2}d_{\gamma a}^{(3)}(z)+d_{\gamma b}^{(3)}(z)\right)O_{\gamma\gamma}(\bq_{1},\bq_{2},\bq_{3})+ {\rm 3\;cyclic}\,\label{eq:G3},
\end{align}
where
\begin{equation}
O_{XY}(\bq_{1},\bq_{2},\bq_{3})=X(\bq_{1},\bq_{2})Y(\bq_{12},\bq_{3})
\end{equation}
(and $\bq_{12}=\bq_{1}+\bq_{2}$). 
In the EdS approximation, the coefficients in the kernels above take the redshift-independent values
\begin{align}
a_{\gamma,\;{\rm EdS}}^{(2)} &= \frac{10}{7}\,,\quad a_{\gamma a,\;{\rm EdS}}^{(3)}  =\frac{2}{3}\,, \quad a_{\gamma b,\;{\rm EdS}}^{(3)}  = \frac{5}{9}\,,\nonumber\\
d_{\gamma,\;{\rm EdS}}^{(2)}  &= \frac{6}{7}\,, \quad d_{\gamma a,\;{\rm EdS}}^{(3)}  = \frac{2}{7}\,, \quad d_{\gamma b,\;{\rm EdS}}^{(3)} = \frac{5}{21}\,,
\label{EdSvals}
\end{align}
while in $\Lambda$CDM or in any other cosmology respecting the Equivalence Principle, they are derived from the perturbative solution of the set of continuity and Euler equations, as described in \cite{DAmico:2021rdb}. After UV-subtraction (as in \cite{Amendola:2022vte}), the redshift space PS at one loop does not depend on $a_{\gamma b,\;{\rm EdS}}^{(3)}$ and $d_{\gamma b,\;{\rm EdS}}^{(3)}$.

For biased tracers we have the following real-space kernels,
\begin{align}
& K_{1}(\bq_{1};z)  = b_1(z)\,,\nonumber\\
& 2 K_{2}(\bq_{1},\bq_{2};z)  =b_{2}(z)+2 b_{1}(z)\beta(\bq_{1},\bq_{2})+c_{\gamma}^{(2)}(z)\gamma(\bq_{1},\bq_{2})\,,\nonumber \\
& 3!\,K_{3}(\bq_{1},\bq_{2},\bq_{3};z)  =\frac{1}{3} c_{0}^{(3)}(z)+2b_{2}\beta(q_{1},q_{2})+ c_{\gamma}^{(3)}(z) \gamma(\bq_{1},\bq_{2})+2b_{1}(z) O_{\beta\beta}(\bq_{1},\bq_{2},\bq_{3})\nonumber\\
 & \quad  +\left[2 c_{\gamma}^{(2)}(z)+c_{\gamma a}^{(3)}(z)-2\left(c_{\gamma b}^{(3)}(z)+ b_{1}(z) \left(a_{\gamma}^{(2)}(z) -1\right)\right)\right]O_{\beta\gamma}(\bq_{1},\bq_{2},\bq_{3})\nonumber\\
 &\quad  +\left(\frac{1}{4} c_{\gamma a}^{(3)}(z)- \frac{1}{2} c_{\gamma b}^{(3)}(z)\right)O_{\gamma\alpha_{a}}(\bq_{1},\bq_{2},\bq_{3})\nonumber\\
 &\quad +\left(c_{\gamma b}^{(3)}(z)-\frac{1}{2} c_{\gamma a}^{(3)}(z)+2 b_{1}(z) \left(a_{\gamma}^{(2)}(z) -1\right)\right)O_{\gamma\beta}(\bq_{1},\bq_{2},\bq_{3})\nonumber\\
 & \quad +\left( \frac{1}{2}c_{\gamma a}^{(3)}(z) +c_{\gamma b}^{(3)}(z)\right) O_{\gamma\gamma}(q_{1},q_{2},q_{3})+ {\rm 3\;cyclic}\,.
\end{align}
The kernels in redshift space for biased tracers are built in terms of the ones for density and velocity as usual,
\begin{align}
    & Z_{1}(\bq; z)   = \left(b_{1}(z)  +\mu_q^{2}\,f(z)\right)\,,\label{z1-1}\\
    & Z_{2}(\bq_1,\bq_2; z)  = K_{2}(\bq_1, \bq_2; z)+f(z)  \mu^{2}_{q_{12}}G_{2}(\bq_1, \bq_2;z)\nonumber \\
     &\quad  + \frac{1}{2} b_{1}(z) f(z)\mu_{q_{12}} q_{12}
     \left(\frac{\mu_{q_1}}{q_1}+\frac{\mu_{q_2}}{q_2}\right)+
    \frac{1}{2} f(z)^2  \mu_{q_{12}}^2 q_{12}^2 \frac{\mu_{q_1}\mu_{q_2}}{q_1 q_2}\,,\nonumber\\
    & Z_{3}(\bq_1,\bq_2,\bq_3; z)  = K_{3}(\bq_1,\bq_2,\bq_3; z) +f(z) \mu_{q_{123}}^2  G_{3}(\bq_1,\bq_2,\bq_3; z) \nonumber \\
    & \quad + \frac{1}{3} f(z) \mu_{q_{123}} q_{123} \bigg[ \frac{\mu_{q_1}}{q_1} K_{2}(\bq_2, \bq_3; z) + \frac{\mu_{q_{23}}}{q_{23}} G_{2}(\bq_2, \bq_3; z)\left( b_1(z)+ f(z) \mu_{q_{123}} q_{123} \frac{\mu_{q_1}}{q_1}\right)\nonumber\\
    &\quad  +b_1(z) f(z)  \mu_{q_{123}} q_{123} \frac{\mu_{q_2}\mu_{q_3}}{q_2 q_3} + {\rm 2\; cyclic}  \bigg] + 
    f(z)^3 \mu_{q_{123}}^3 q_{123}^3 \frac{\mu_{q_1}\mu_{q_2}\mu_{q_3}}{q_1 q_2 q_3}\,.
    \label{ZK}
\end{align}
Once UV-subtracted, the redshift space PS at one loop depends only on the four tracer-dependent parameters
\begin{equation}
    b_1(z),\; b_2(z),\;c_{\gamma}^{(2)}(z),\;c_{\gamma a}^{(3)}(z),
    \label{TDpars}
\end{equation}
and on the four  cosmology-dependent (and tracer-independent) ones,
\begin{equation}
f(z),\;d_{\gamma}^{(2)}(z),\;d_{\gamma a}^{(3)}(z),\;a_{\gamma}^{(2)}(z).
\label{CDpars}
\end{equation}
 Notice that, while $a_{\gamma}^{(2)}(z)$ is formally second order, it appears in the redshift space kernels only through the $h(z)$ dependence, which, for biased tracers, appears at third order.

In a fixed cosmology, namely, fixing the $a$'s and $d$'s coefficients of the matter and velocity kernels, the last two tracer dependent coefficients in (\ref{TDpars}) can be expressed in terms of the $b_{{\mathcal{G}}_{2}}$ and $b_{\Gamma_{3}}$ parameters of the standard bias expansion (using the notation of \cite{Ivanov:2019pdj}) as
\begin{align}    
    c_{\gamma}^{(2)} & =b_{1}a_{\gamma}^{(2)}-2b_{{\mathcal{G}}_{2}}\,,\nonumber\\
    c_{\gamma a}^{(3)} & =b_{1}a_{\gamma a}^{(3)}-2b_{{\mathcal{G}}_{2}}a_{\gamma}^{(2)}-2b_{\Gamma_{3}}(a_{\gamma}^{(2)}-d_{\gamma}^{(2)})\,,
    \label{biasvoc}
\end{align}
 or, inversely, 
\begin{align}
    \label{biasvoc2}
    b_{{\mathcal{G}}_{2}} & = \frac{1}{2} \left(b_{1}a_{\gamma}^{(2)} -  c_{\gamma}^{(2)}  \right)\,,\nonumber\\
    b_{\Gamma_{3}} & = \frac{b_1 a_{\gamma a}^{(3)} + a_{\gamma}^{(2)}\left(c_{\gamma}^{(2)}- b_1 a_{\gamma}^{(2)} \right)  - c_{\gamma a}^{(3)} }{2\left(a_{\gamma}^{(2)}-d_{\gamma}^{(2)} \right)}\,.
\end{align}
In this paper,  we  use $a_{\gamma a}^{(3)}$ as free parameter, instead of $c_{\gamma a}^{(3)}$, and take into account its degeneracy with $b_{\Gamma_{3}}$ by fixing the latter. This procedure is not tracer independent, therefore the only tracer-independent parameters  are the ones in (\ref{CDpars}).

When a specific cosmological model is considered, it is possible to study the time evolution of the bootstrap functions. For the growth function $f = d\ln{D}/d\ln{a}$ the most general equation in cosmologies that satisfy the equivalence principle is~\cite{DAmico:2021rdb,Piga:2022mge}
\begin{equation}
    \frac{d f}{d\ln{a}} + \left(2 + f  + \frac{d\ln{E(a)}}{d\ln{a}}\right)f - \frac{3}{2}\mu(a)\Omega_m(a) = 0\,,
\end{equation}
where $\mu(a)$ is a time dependent function that accounts for possible extensions in the gravitational interaction through the Poisson equation. The background functions are 
\begin{equation}
    E(a) = \frac{H(a)}{H_0} = \sqrt{\Omega_{m0} a^{-3} + \Omega_{{\rm DE}0}a^{-3 w} + \Omega_{k0} a^{-2}}\,,\quad \quad \Omega_m(a) = \frac{\Omega_{m0}a^{-3}}{E(a)^2}\,.
\end{equation}
Beyond-standard models enter in the evolution of $f$ through $\mu$ and the background evolution, for example for $w_0$-$w_a$ models one has $w = w_0 + w_a(1-a)$ for the dark energy evolution equation. For the second order bootstrap functions the equations are 
\begin{align}
&\frac{d\, a_{\gamma}^{(2)} }{d\ln{a}} = \left(2 - 2 a_{\gamma}^{(2)} + d_{\gamma}^{(2)} \right)f\,, \\
& \frac{d\, d_{\gamma}^{(2)}}{d\ln{a}} = \left(\frac{3}{2}\frac{\Omega_m}{f^2} \mu \bigg(a_{\gamma}^{(2)} - d_{\gamma}^{(2)} \bigg) - d_{\gamma}^{(2)}+ 2 \frac{\mu_{2}}{f^2}\bigg(\frac{3}{2}\Omega_m \bigg)^2\right)f\,, \nonumber
\end{align}
where $\mu$ and $\mu_2$ are, respectively, the functions that describes linear and non-linear modifications to the Poisson equation (in GR, $\mu = 1$ and $\mu_2 = 0$).

\section{Full triangle plots}
\label{app:full_tri}
\begin{figure}
    \centering
    \includegraphics[width=0.95\textwidth]{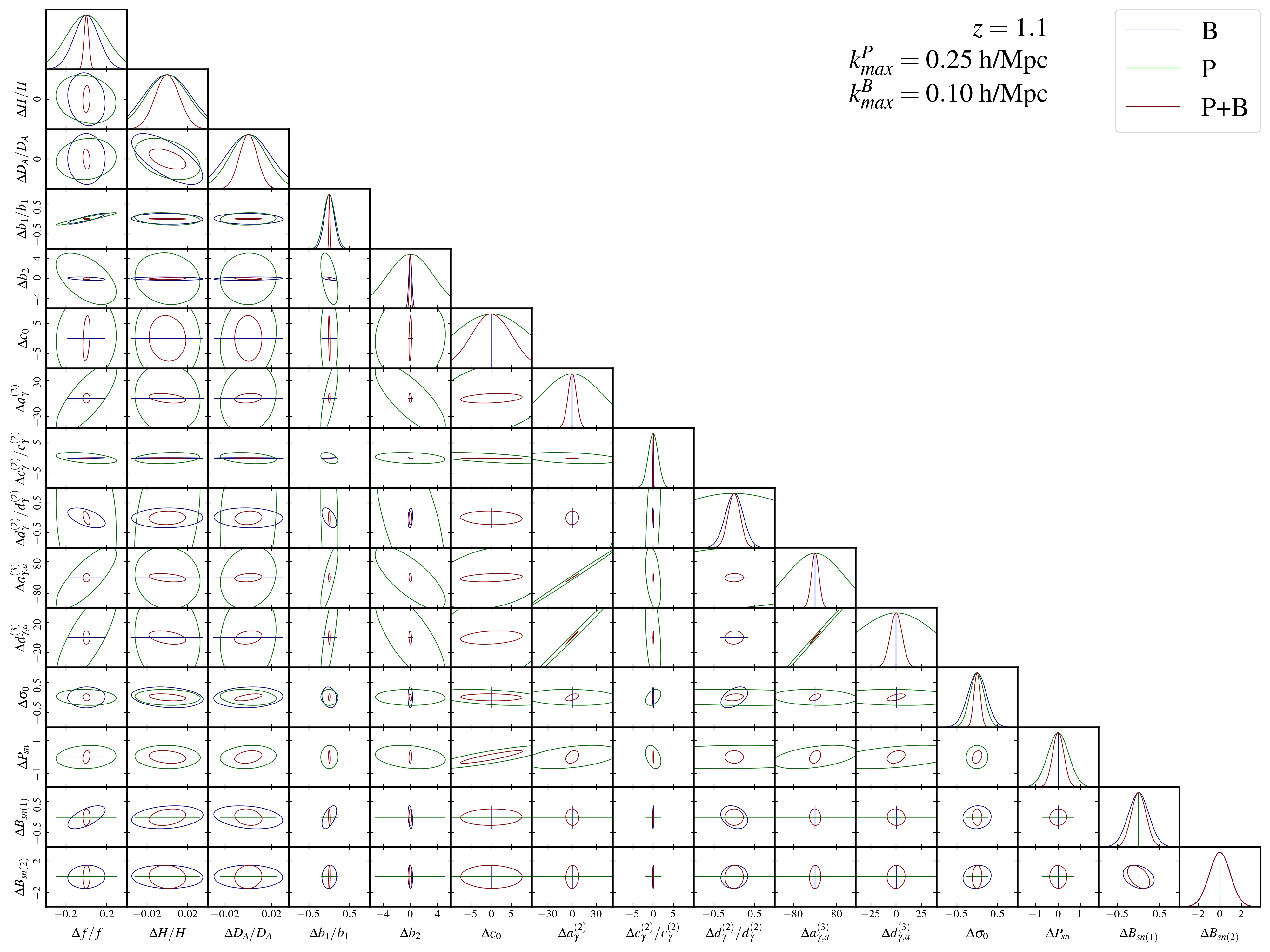}
    \caption{ 1-$\sigma$ triangle plot for all the parameters varied in this Fisher analysis in the aggressive case.}
    \label{fig:fish_tot_aggr}
\end{figure}

For completeness we show in Figure~\ref{fig:fish_tot_aggr} the triangle plot for all the $z$-dependent parameters  in the aggressive case. We show only the 68$\%$ confidence regions to avoid clutter.

\bibliographystyle{JHEP2015}
\bibliography{references,scaling_bib,curvature}
 \label{lastpage} 
\end{document}